\documentclass[12pt,english,a4paper,prd,nofootinbib,
preprint,
showpacs,
preprintnumbers,
floatfix]{revtex4}
\usepackage{mathptmx}
\usepackage{helvet}
\usepackage{courier}
\usepackage[T1]{fontenc}
\usepackage[a4paper]{geometry}
\usepackage{array}
\usepackage{amsmath}
\usepackage{amssymb}
\usepackage{graphicx}
\usepackage{babel}
\usepackage{color}

\makeatletter

\providecommand{\tabularnewline}{\\}
\newcommand{\lyxdot}{.}

\@ifundefined{textcolor}{}
{%
 \definecolor{BLACK}{gray}{0}
 \definecolor{WHITE}{gray}{1}
 \definecolor{RED}{rgb}{1,0,0}
 \definecolor{GREEN}{rgb}{0,1,0}
 \definecolor{BLUE}{rgb}{0,0,1}
 \definecolor{CYAN}{cmyk}{1,0,0,0}
 \definecolor{MAGENTA}{cmyk}{0,1,0,0}
 \definecolor{YELLOW}{cmyk}{0,0,1,0}
}

\makeatother

\begin{document}

\global\long\def\ra{\rightarrow}
\global\long\def\alphas{\alpha_{s}}
\global\long\def\ve{\varepsilon}
\global\long\def\gae{\gamma_{E}}
\global\long\def\AcalFB{{\cal A}_{FB}}

\newcommand\gsim{
\mathrel{\rlap{\raise.4ex\hbox{$>$}} {\lower.6ex\hbox{$\sim$}}}}  
\newcommand\lsim{
\mathrel{\rlap{\raise.4ex\hbox{$<$}} {\lower.6ex\hbox{$\sim$}}}}  
\newcommand\etmiss{E_{T}\hspace{-13pt}/\hspace{8pt}}
\newcommand{\mHQ}{m_{\mathcal Q}} 
\newcommand{\SLASH}[2]{\makebox[#2ex][l]{$#1$}/} 
\newcommand{\Dslash}{\SLASH{D}{.5}\,} 
\newcommand{\kslash}{\SLASH{k}{.15}} 
\newcommand{\nslash}{\SLASH{n}{.15}} 
\newcommand{\pslash}{\SLASH{p}{.2}} 
\newcommand{\qslash}{\SLASH{q}{.08}} 
\newcommand{\syslash}{\SLASH{s}{.06}} 
\newcommand{\aslash}{\SLASH{a}{.06}} 
\newcommand{\bslash}{\SLASH{b}{.06}} 
\newcommand{\parsl}{\SLASH{\partial}{.1}}
\newcommand{\Mcal}{\mathcal{M}}

\begin{flushright}
MITP/14-001\\
TTK-14-10
\par\end{flushright}

\vspace{4mm}
 
\begin{center}
\textbf{\LARGE Top Quark Charge Asymmetry: }
\par\end{center}{\LARGE \par}

\begin{center}
\textbf{\LARGE Searching for Light Axigluons in }
\par\end{center}{\LARGE \par}

\begin{center}
\textbf{\LARGE $t\bar{t}+jet$ Production at the LHC }
\vspace{12mm}
\par\end{center}

\begin{center}
\textbf{\large Stefan Alte}$^{\, a,}$\textbf{\large }%
\footnote{\texttt{\small salte@students.uni-mainz.de}%
}\textbf{\large , Stefan Berge}$^{\, a,b,}$\textbf{\large }%
\footnote{\texttt{\small berge@physik.rwth-aachen.de}%
}\textbf{\large , Hubert Spiesberger}$^{\, a,}$\textbf{\large }%
\footnote{\texttt{\small spiesber@uni-mainz.de}%
}
\par\end{center}{\large \par}

\begin{center}
$^{a}$ PRISMA Cluster of Excellence, Institut f\"ur Physik (WA THEP),
\\
Johannes Gutenberg-Universit\"at, 55099 Mainz, Germany
\par\end{center}

\begin{center}
$^{b}$ Institut f\"ur Theoretische Teilchenphysik und Kosmologie, 
\\
RWTH Aachen, 52056 Aachen, Germany
\par\end{center}

\begin{center}
\vspace{12mm}
\textbf{Abstract}
\par\end{center}

We investigate the discovery potential of light color-octet bosons
in the mass range of $100 - 400$ GeV in exclusive top-pair plus
jet production at the LHC, $pp\rightarrow t\bar{t}+jet$. We study the
impact of such bosons on the incline, the energy and the rapidity
asymmetries. We show that light axigluons with large couplings to 
quarks can be discovered at the LHC with a luminosity of a few 
fb$^{-1}$. Almost all of the considered axigluon parameter space 
can be probed using the already available 2011/2012 LHC data. In 
a small-coupling scenario, axigluons could be discovered using the 
charge asymmetry with 65~fb$^{-1}$ at the LHC and a center of mass 
energy of $14$~TeV. We furthermore show that $t\bar{t}+jet$ production 
could reveal the existence of scenarios where axigluons couple with 
a different strength to up- and down-type quarks.

\thispagestyle{empty}

\newpage{}

\section{Introduction}

For inclusive quark pair production, $q\bar{q}\to Q\bar{Q}$, at 
hadron colliders, QCD predicts a charge asymmetry arising at NLO 
from virtual and real gluon radiation~\cite{Brown:1979dd}. This 
charge asymmetry has been studied for top quarks at the Tevatron 
and LHC experiments in~\cite{Kuhn:1998jr, Kuhn:1998kw}. At the 
Tevatron the charge asymmetry has been measured as a forward-backward 
asymmetry at the level of the top quarks~\cite{Aaltonen:2012it, 
Abazov:2014cca} and at the level of the decay products in the 
semi-leptonic~\cite{Aaltonen:2013vaf, Abazov:2014oea} and double 
leptonic decay~\cite{Aaltonen:2014eva, Abazov:2013wxa}. While CDF 
found deviations from the NLO Standard Model prediction~\cite{Kuhn:2011ri, 
Bernreuther:2012sx, Hollik:2011ps} in particular in the region of high 
invariant mass of the $t\bar{t}$ pair, $M_{t\bar{t}}$, and at large 
rapidity differences, $\Delta y$, the most recent analyses from D0 
agree with the SM prediction if all analysis channels are combined. 
At the LHC, measurements of the top-quark charge asymmetry in 
semi-leptonic~\cite{Aad:2013cea, Chatrchyan:2012cxa, CMS:2013nfa} 
and double-leptonic decays~\cite{ATLAS:2012sla, Chatrchyan:2014yta} 
agree with the SM prediction; however, the errors are still very 
large \cite{CMS-PAS-TOP-14-006}. 

For the case of $t\bar{t}$ production in association with a jet
at hadron colliders, a charge asymmetry is already generated at 
the leading order (LO)~\cite{Halzen:1987xd} and can be measured
as an asymmetry in the rapidity difference of the top and antitop 
quark. Also for $t\bar{t}+jet$ production, the next-to-leading order 
(NLO) QCD corrections of the rapidity asymmetry have been 
investigated~\cite{Dittmaier:2007wz, Dittmaier:2008uj, Melnikov:2010iu, 
Ahrens:2011uf}. Furthermore, the effects of the top-quark decay 
and of parton showers have been discussed in~\cite{Alioli:2011as, 
Melnikov:2011qx, Kardos:2011qa}. In Ref.~\cite{Hoeche:2013mua} it 
was found that data for the dependence of the charge asymmetry on 
the transverse momentum of the $t\bar{t}$-system agree with the 
SM predictions obtained from a combination of $t\bar{t}$ and 
$t\bar{t}+jet$ calculations at NLO, merged with parton showers. 
The rapidity and mass dependences, however, still show some 
discrepancies. 

To explain the discrepancies observed in the Tevatron measurements 
of the forward-backward asymmetry in inclusive $t\bar{t}$ production, 
a number of different models have been suggested (current reviews 
include~\cite{Westhoff:2013ixa, Aguilar-Saavedra:2013rza, 
Berger:2013et, AguilarSaavedra:2011vw, Kamenik:2011wt}). One of them 
is the axigluon model introduced in~\cite{Frampton:1987dn} and first 
studied in \cite{Bagger:1987fz}. It is based on the assumption that 
SM QCD emerges after spontaneous symmetry breakdown from a gauge 
theory based on the chiral group $SU(3)_L \times SU(3)_R$. It predicts 
the existence of heavy partners of the SM gluons which have axial-vector 
couplings to the SM quarks. These axigluons would lead to an enhancement 
of the forward-backward asymmetry in $t\bar{t}$ production.

Heavy axigluons with masses larger than $\sim 1$~TeV are already 
strongly constrained by collider measurements~\cite{Chivukula:2010fk, 
Diaz:2013tfa, Haisch:2011up}. In Ref.~\cite{Xiao:2010ph} it was shown 
that the deviation from the SM prediction to the measured Tevatron 
top forward-backward asymmetry could also be due to light axigluons. 
This idea has been discussed in different scenarios for axigluons 
with masses in the range of $50-1000$~GeV~\cite{Tavares:2011zg, 
Barcelo:2011vk, Alvarez:2011hi, AguilarSaavedra:2011ci, Krnjaic:2011ub}, 
and usually with a large axigluon decay width to prevent strong bounds 
from measurements of $t\bar{t}$ invariant mass spectra. Extensive 
studies of the phenomenology of light axigluons have been performed, 
including~\cite{Aguilar-Saavedra:2014vta, Tianjun:2013joa, 
Carmona:2014gra, Aguilar-Saavedra:2014yea, Jung:2014gfa, Ipek:2013zi, 
Falkowski:2012cu, Baumgart:2013yra, Grinstein:2013mia, 
Baumgart:2012ay}. In particular, in Ref.~\cite{Aguilar-Saavedra:2014nja} 
it could be demonstrated that a fit of top quark measurements at 
the Tevatron and the LHC performed within the SM is improved 
considerably if a light axigluon is included in the theory 
(compare also~\cite{Gresham:2012kv, Yue:2014hba, Aaltonen:2013hya, 
Dobrescu:2013cmh, Chen:1991tx} for model constraints from
other  measurements).

In the present paper we will therefore investigate the impact of such 
light axigluons on the charge asymmetries in $t\bar{t}+jet$ 
production for the LHC. We choose the parameter space as suggested 
in~\cite{Gross:2012bz} with axigluon masses in the range of $100 - 
400$~GeV. Recently, in Ref.\ \cite{Berge:2013xsa} it was demonstrated, 
that the anti-symmetric part of the cross section of the LO partonic 
$t\bar{t}+jet$ process can be separated into two independent terms, 
allowing for different possibilities to define a charge asymmetry. 
We will base our investigations on the so-called incline and energy 
asymmetries, as well as on the conventional rapidity asymmetry. We 
will discuss specific features of the three possible definitions of 
the charge asymmetry and demonstrate that they are very sensitive to 
light axigluons. We furthermore will discuss a flipped (down-type 
non-universal) scenario~\cite{Drobnak:2012cz}, where the axigluon 
coupling to up-type quarks is different from the axigluon coupling 
to down-type quarks.

This paper is organized as follows: In Sec.~\ref{sec:Theory} we define 
the energy and the incline asymmetry~\cite{Berge:2013xsa} as well as 
the conventional rapidity asymmetry. Some properties of the axigluon 
model and its parameter space are stated as far as they are important 
for our work. We also discuss general features of the cross section 
and its separation into parts which are symmetric and anti-symmetric 
with respect to the jet angle. Firstly, we recapitulate these properties 
for the SM and then work out which features of the jet-angle dependence 
of the asymmetries are distinctive of the axigluon model. Subsequently, 
in Sec.~\ref{sub:Parton-level-distributions} we describe in detail 
numerical results for the asymmetries at the parton level for the 
$q\bar{q}$ and the $qg$ initial state and then, in 
Sec.~\ref{sec:hadron-level-results}, at the hadron level for the LHC 
with a center of mass energy of $14$~TeV. Based on these results we 
can determine the discovery potential at the LHC and its dependence on 
the axigluon parameters and the available luminosity. We also consider 
$t\bar{t}+jet$ production at the LHC with a center of mass energy of 
$8$~TeV in Sec.~\ref{Sec:LHC8}. Finally, in Sec.~\ref{Sec:FlippedScenario}, 
we discuss whether and to what extent the flipped scenario, suggested 
in Ref.~\cite{Drobnak:2012cz}, can be revealed in data for the charge 
asymmetry in  $t\bar{t}+jet$ production. Our conclusions are presented 
in Sec.~\ref{Sec:Conclusion}.

\section{Theory
\label{sec:Theory}
}

$t\bar{t}+jet$ final states are created by partonic processes 
$p_{1}p_{2}\to t\bar{t}p_{3}$. We denote the momenta of the 
(anti-)top-quarks by $k_t$ and $k_{\bar{t}}$, respectively, 
and the differential cross section by $\text{d}\hat{\sigma}_{\, 
t\bar{t}j} = \text{d}\hat{\sigma}(p_{1}p_{2} \to t(k_{t})\, 
\bar{t}(k_{\bar{t}})\, p_{3})$. The differential charge asymmetry 
is obtained by subtracting from $\text{d}\hat{\sigma}_{\,t\bar{t}j}$
the amplitude where the top-quark is exchanged with its anti-particle. 
We define the anti-symmetric part of the differential cross section, 
$\text{d}\hat{\sigma}_{A}$, by
\begin{eqnarray}
\text{d}\hat{\sigma}_{A} 
& = & 
\text{d}\hat{\sigma}_{\, t\bar{t}j} - \text{d}\hat{\sigma}_{\,\bar{t}tj} 
\label{eq:sigmaapartdiff}
\end{eqnarray}
where $\text{d}\hat{\sigma}_{\,\bar{t}tj} = 
\text{d}\hat{\sigma}(p_{1}p_{2} \to \bar{t}(k_{t})\, t(k_{\bar{t}})\, 
p_{3})$. 

At LO, the phase space for $t\bar{t}+jet$ production is determined
by four variables. We use the parameterization given in Ref.\ 
\cite{Berge:2013xsa}. Explicit expressions for the differential 
cross sections can be found in the same reference \cite{Berge:2013xsa}. 
With these definitions we can construct differential asymmetries with 
fixed, suitably chosen kinematical variables. In the present paper we 
will study three specific definitions of the charge asymmetry: the 
energy asymmetry, the rapidity asymmetry and the incline asymmetry. 
They are based on the kinematic variables (i) $\Delta E = E_{t} - 
E_{\bar{t}}$, the difference of the energies of the top- and 
anti-top-quark at the parton level, (ii) $\Delta y = y_{t} - 
y_{\bar{t}}$, the difference of the rapidities of the top- and 
anti-top-quark\footnote{
  Below, when we discuss asymmetries at the hadron level, we will 
  use $\Delta |y| = |y_{t}| - |y_{\bar{t}}|$ instead of $\Delta y$.
}
and (iii) $\cos\varphi$, the incline angle of the $t\bar{t}j$-plane. 
We denote these variables collectively by $C$, $C \in \{\Delta E,\, 
\Delta y,\, \cos\varphi\}$. Then, using 
\begin{eqnarray}
\text{d}\hat{\sigma}_{A}^{C} 
& = 
\text{d}\hat{\sigma}_{A}(C>0)
\,\,=\,\, 
& 
\text{d}\hat{\sigma}_{\, t\bar{t}j}(C>0) - 
\text{d}\hat{\sigma}_{\, t\bar{t}j}(C<0) 
\, ,  
\label{eq:sigma_A_partdiff_C} 
\end{eqnarray}
we define normalized partonic asymmetries by 
\begin{eqnarray}
\hat{A}^{C}(X) 
& = & 
\frac{\text{d}\hat{\sigma}_{A}^{C}/\text{d}X}%
 {\text{d}\hat{\sigma}_{S}/\text{d}X}  
\label{eq:parton_diffA}
\end{eqnarray}
where $X$ is a suitably chosen kinematic variable and we have 
also introduced the symmetric differential cross section\footnote{
   This is the symmetric expression $\text{d}\hat{\sigma}_{S} = 
   \text{d}\hat{\sigma}_{\, t\bar{t}j}(C>0) + 
   \text{d}\hat{\sigma}_{\, t\bar{t}j}(C<0)$, defined in analogy 
   with Eq.\ (\ref{eq:sigma_A_partdiff_C}), which is independent 
   of $C$ and simply equal to 
   $\text{d}\hat{\sigma}_{\, t\bar{t}j}$.
   } 
$\text{d}\hat{\sigma}_{S} = \text{d}\hat{\sigma}_{\, t\bar{t}j}$. 
Integrating over the full phase space and defining the partonic 
anti-symmetric and symmetric cross sections, $\hat{\sigma}_{A}^{C}$ 
and $\hat{\sigma}_{S}$, we also define integrated asymmetries
\begin{eqnarray}
\hat{A}^{C}_{\text{int}} 
& = & 
\frac{\hat{\sigma}_{A}^{C}}{\hat{\sigma}_{S}} \, .
\label{eq:parton_chargeasymmetry}
\end{eqnarray}
Note that $\hat{\sigma}_{S}$, i.e.\ $\text{d}\hat{\sigma}_{S}$ 
integrated over the full phase space, is equal to the total partonic cross 
section for $t\bar{t}j$ production: $\hat{\sigma}_{S} = 
\hat{\sigma}_{t\bar{t}j}$. 

The charge asymmetry at the hadron level is calculated from the 
parton-level cross sections after a convolution with the parton 
distribution functions (PDFs) $f_{p_{i}/N_{i}}(x_{i},\mu_{f})$
at the factorization scale $\mu_{f}$: 
\begin{align}
A^{C}(X) &=  
\frac{\text{d}\sigma_{A}^{C}/\text{d}X}{\text{d}\sigma_{S}/\text{d}X} 
\, , 
\nonumber
\\
\text{d}\sigma_{A}^{C} 
&= 
\sum_{p_{1}p_{2}} \int\text{d}x_{1}\text{d}x_{2} \, 
f_{p_{1}/N_{1}}(x_{1},\mu_{f})\, f_{p_{2}/N_{2}}(x_{2},\mu_{f}) \, 
\text{d}\hat{\sigma}_{A}^{C,p_{1}p_{2}}(\hat{s},\mu_{f}) 
\, , 
\nonumber
\\
\text{d}\sigma_{S} 
&= 
\sum_{p_{1}p_{2}} \int\text{d}x_{1}\text{d}x_{2} \, 
f_{p_{1}/N_{1}}(x_{1},\mu_{f})\, f_{p_{2}/N_{2}}(x_{2},\mu_{f}) \, 
\text{d}\hat{\sigma}_{S}^{p_{1}p_{2}}(\hat{s},\mu_{f})  
\label{eq:chargeasymmetry-1}
\end{align} 
and similar equations for integrated cross sections and asymmetries. 
Here, $x_{i}$ are the momentum fractions of the partons $p_{i}$
inside the nucleons $N_{i}$, $\hat{s}$ is the squared partonic 
center-of-mass (CM) energy related to the CM energy of the colliding 
nucleons, $S$, by $\hat{s} = x_{1}x_{2}S$. A non-zero asymmetry 
is generated in the channels $p_{1}p_{2} = q\bar{q}$, $\bar{q}q$, 
$qg$, $gq$, $\bar{q}g$, $g\bar{q}$, but not in the $gg$-channel. 

At the Tevatron, the top-quark pair production is dominated by the 
$q\bar q$ channel, while at the LHC, the quark-gluon initial state 
will also be important. The boost of the partonic CM frame with
respect to the laboratory frame will be expressed in terms of the 
rapidity of the top-anti-top-jet system in the laboratory 
frame\footnote{This relation is valid at leading order.}, 
$y_{t\bar{t}j} = \ln(x_{1}/x_{2})/2$.

Apart from Sec.~\ref{Sec:FlippedScenario} we will consider models 
which contain light axigluons with couplings to the SM quarks of 
purely axial-vector type. The part of the Lagrangian relevant for 
top-anti-top production is 
\begin{eqnarray}
\mathcal{L}_{axi} 
& = & 
g_{A}^{i} \, 
\bar{q}_{i} \, \gamma^{\mu}\gamma_{5} T^{a} q_{i} \, 
G_{\mu}^{a} 
- g_{s}\, f_{abc} \, 
\big[(\partial_{\mu}G_{\nu}^{a} - \partial_{\nu}G_{\mu}^{a}) \, 
G^{b\mu}g^{c\nu} + G^{a\mu}G^{b\nu}(\partial_{\mu}g_{\nu}^{c}) 
\big] 
\, . 
\label{eq:Axi_lagrangian}
\end{eqnarray}

Here $g_{\mu}^{a}$ denotes the SM gluon field, $G_{\mu}^{a}$ the
massive axigluon field and $q_{i}$ the SM quark fields. $T^{a}$
are the generators and $f^{abc}$ the structure constants of the
$SU(3)$ gauge group. $g_{s}$ is the strong coupling constant of
the SM and $g_{A}^{i}$ the axial-vector couplings of the massive 
gluon to quarks with flavor $i$ in their weak eigenstates. In the 
absence of vector couplings of the axigluons, only the product of 
$g_{A}^{q}$ and $g_{A}^{t}$ (where $q=\{u,d,c,s,b\}$) appears in 
the cross section for $t\bar{t}+jet$ production~\cite{Berge:2012rc}.
For convenience we also define the coupling factor 
\cite{Gross:2012bz}
\begin{eqnarray}
\alpha_{A} 
& = & 
\frac{g_{A}^{q}g_{A}^{t}}{4\pi}  
\, . 
\label{eq:def_alpha_A}
\end{eqnarray}
In accordance with Ref.~\cite{Gross:2012bz}, the mass $m_A$ of the 
axigluons is chosen to vary in the range $100 \le m_A \le 400$~GeV 
and the width of the axigluon should be large: $\Gamma_{A} = 0.1\cdot 
m_A$. As we will discuss below at the end of 
Sec.~\ref{sub:Parton-level-distributions}, larger values of the 
axigluon width would only have little effect on the charge asymmetry. 
In our numerical calculations we will highlight two coupling scenarios: 
A small-coupling scenario, defined by $m_{A} = 300$~GeV, $\alpha_{A} 
= 0.005$ and $\Gamma_{A} = 0.1 \cdot m_{A}$, resulting in the smallest 
deviations of the charge asymmetries with respect to the SM prediction; 
and, secondly, a large-coupling scenario defined by $m_{A} = 400$~GeV, 
$\alpha_{A} = 0.032$ and $\Gamma_{A} = 0.1 \cdot m_{A}$. The 
large-coupling scenario was found to lead to the largest possible 
deviations of the total charge asymmetry from its SM predictions 
within the considered axigluon parameter space.

\begin{figure}[!t]
\begin{centering}
~~~
\includegraphics[height=2.4cm]{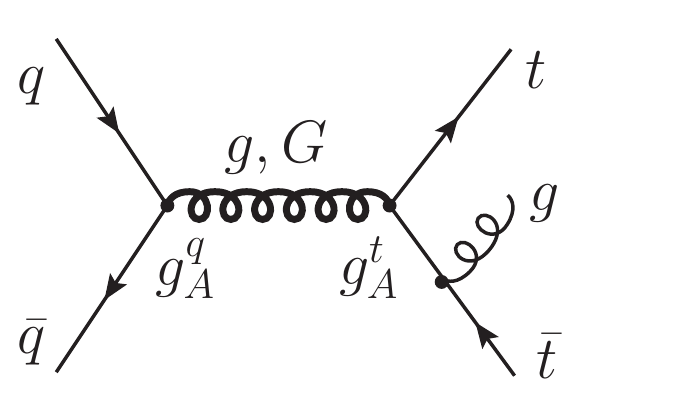}
\hspace*{.6cm}
\includegraphics[height=2.4cm]{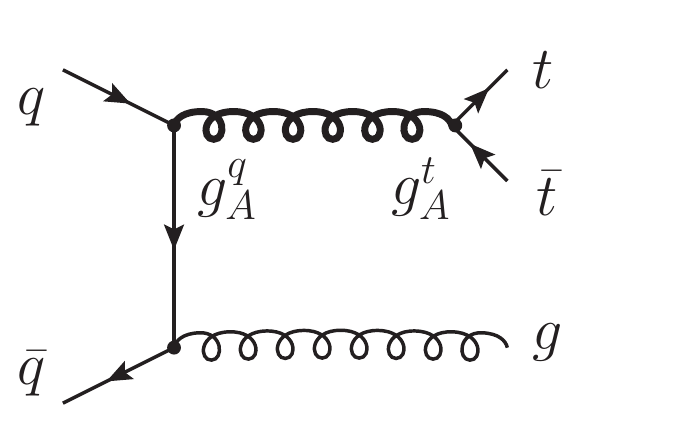}
\hspace*{.6cm}
\includegraphics[height=2.4cm]{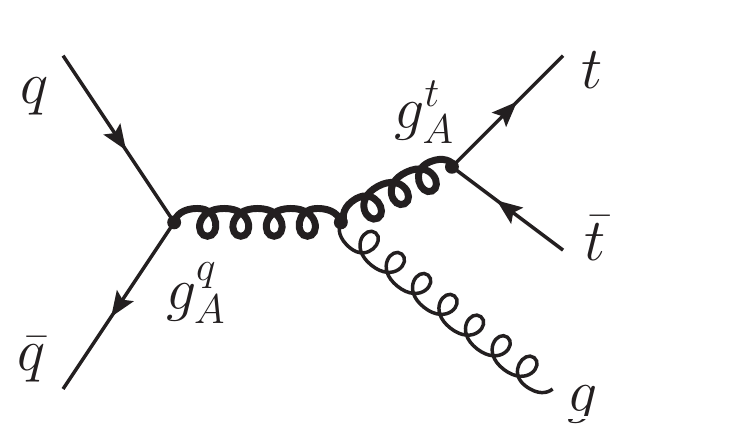}

{\footnotesize (a)}
\hspace{4.4cm}
{\footnotesize (b)}
\hspace{4.2cm}
{\footnotesize (c)}
\end{centering}
\caption{
\label{fig:ttj_axi} 
Feynman diagrams contributing to the partonic process  $q\bar{q}
\rightarrow t\bar{t}g$. Thin curly lines denote SM gluons while
thick curly lines denote either a SM gluon ($g$) or an axigluon 
($G$). Axial-vector couplings of quarks to axigluons are indicated 
by $g_A^q$ and axial-vector couplings of top quarks to axigluons
by $g_A^t$. Two more Feynman diagrams with the gluon attached to 
$t$ instead of $\bar{t}$ ($q$ instead of $\bar{q}$ in (b)) are not 
shown. 
} 
\end{figure}

The Feynman diagrams for the leading contributions in the partonic 
sub-process $q\bar{q} \to t\bar{t}g$ are shown in 
Fig.~\ref{fig:ttj_axi}.
The thin curly outgoing lines represent 
the QCD gluon while the thick curly lines can stand for either a 
QCD gluon or an axigluon. In order to analyze the structure of 
the various contributions to the cross section it will be helpful 
to separate the modulus squared of the sum of the Feynman diagrams 
into the sum of products $M_i^\ast M_j$ where $M_i$ denotes one 
of the Feynman diagrams in Fig.~\ref{fig:ttj_axi}, $i,j = a$, $b$, 
$c$. Each product can be further split into a pure SM contribution, 
denoted $\hat{\sigma}_{S/A}^{g}$; a term without any virtual SM 
gluon, denoted $\hat{\sigma}_{S/A}^{G}$; and a rest, i.e.\ interference 
terms of a SM-like diagram with a diagram containing one or two 
axigluons, denoted $\hat{\sigma}_{S/A}^{gG}$. In 
Tab.~\ref{tab:sigmaa_axi} we have collected the various terms and 
show how each contribution depends on the coupling constants. 

\begin{table}[h]
\begin{centering}
\global\long\def\arraystretch{1.1}
\begin{tabular}{|l||>{\centering}p{1cm}|>{\centering}p{1.1cm}|>{\centering}p{1.4cm}||>{\centering}p{1cm}|>{\centering}p{1.1cm}|>{\centering}p{2.6cm}|}
\hline 
& $\hat{\sigma}_{S}^{g}$ & $\hat{\sigma}_{S}^{gG}$ & $\hat{\sigma}_{S}^{G}$  & $\hat{\sigma}_{A}^{g}$ & $\hat{\sigma}_{A}^{gG}$ & $\hat{\sigma}_{A}^{G}$\tabularnewline
\hline 
\hline 
~~$M_a^\ast M_a,\,M_b^\ast M_b,\,M_c^\ast M_c$~~ & $1$ & $-$ & $(g_{A}^{q}g_{A}^{t})^{2}$ & $-$ & $g_{A}^{q}g_{A}^{t}$  & $-$ 
\tabularnewline
~~$M_a^\ast M_b:\, f_{abc}^{2}$ & $1$ & $-$ & $(g_{A}^{q}g_{A}^{t})^{2}$ & $-$ & $g_{A}^{q}g_{A}^{t}$  & $-$ 
\tabularnewline
~~$M_a^\ast M_b:\, d_{abc}^{2}$ & $-$ & $g_{A}^{q}g_{A}^{t}$ & $-$ & $1$ & $-$ & $(g_{V,A}^{q})^{2}(g_{V,A}^{t})^{2}$ 
\tabularnewline
~~$M_a^\ast M_c,\,M_b^\ast M_c$~~ & $1$ & $-$ & $(g_{A}^{q}g_{A}^{t})^{2}$ & $-$ & $g_{A}^{q}g_{A}^{t}$ & $-$ 
\tabularnewline
\hline 
\end{tabular}
\par\end{centering}
\caption{ 
\label{tab:sigmaa_axi} 
Contributions to the $t\bar{t}+j$ cross section originating from 
SM Feynman diagrams ($\hat{\sigma}_{S/A}^{g}$, columns 2 and 5), 
from SM gluon-axigluon interference ($\hat{\sigma}_{S/A}^{gG}$, 
columns 3 and 6), and 
from diagrams containing no virtual SM gluons
($\hat{\sigma}_{S/A}^{G}$, columns 4 and 7).
Each entry contains the product of coupling constants to axigluons 
contained in the product of matrix elements shown in column 1. 
The labels $a$, $b$, and $c$ correspond to the diagrams shown in 
Fig.~\ref{fig:ttj_axi}.
} 
\end{table}

We discuss the various contributions listed in 
Tab.~\ref{tab:sigmaa_axi} in turn. The anti-symmetric part of the 
pure SM contribution, $\hat{\sigma}_{A}^{g}$, receives a contribution 
only from the interference of diagrams (a) and (b) and this part is 
proportional to $d_{abc}^{\,2}$, the symmetric structure constants 
of $SU(3)$ \cite{Kuhn:1998kw}. This can be seen as follows. Firstly, 
one observes that the Lorentz structure of this interference term 
is anti-symmetric with respect to charge conjugation\footnote{This 
   can be seen by relating the interference term via the optical 
   theorem to a loop diagram and using the Furry theorem.}. 
On the other hand, the color coefficient is $1/(16N_{c}^{2}) \cdot 
(d_{abc}^{2}+f_{abc}^{2})$ for $t\bar{t}j$ and $1/(16N_{c}^{2}) 
\cdot(d_{abc}^{2}-f_{abc}^{2})$ if $t$ and $\bar{t}$ are interchanged. 
Therefore, the term proportional to $d_{abc}^{2}$ contributes to 
$\hat{\sigma}_{A}^{g}$ while the term proportional to $f_{abc}^{2}$ 
contributes to $\hat{\sigma}_{S}^{g}$. With similar arguments one 
can show that all other contributions in the first column of 
Tab.~\ref{tab:sigmaa_axi} contribute only to $\hat{\sigma}_{S}^{g}$.

The SM-axigluon interference terms $\hat{\sigma}_{A}^{gG}$ and 
$\hat{\sigma}_{S}^{gG}$ contain an axial coupling $\gamma_{\mu} 
\gamma_{5}$ from the quark-anti-quark-axigluon vertex (see 
Eq.~\eqref{eq:Axi_lagrangian}) which leads to a different Lorentz 
structure. Therefore, interference terms which were anti-symmetric 
under charge conjugation for a pure QCD diagram are now symmetric 
and vice versa. For $\hat{\sigma}_{A}^{G}$ and $\hat{\sigma}_{S}^{G}$, 
due to the presence of two $\gamma_{\mu}\gamma_{5}$ couplings in 
the top-quark lines, the products of Feynman diagrams listed in the 
first column of Tab.~\ref{tab:sigmaa_axi} contribute in the same 
way to the symmetric and anti-symmetric parts of the cross section 
as the pure SM contributions.

\begin{figure}[t]
\includegraphics[height=5.5cm]{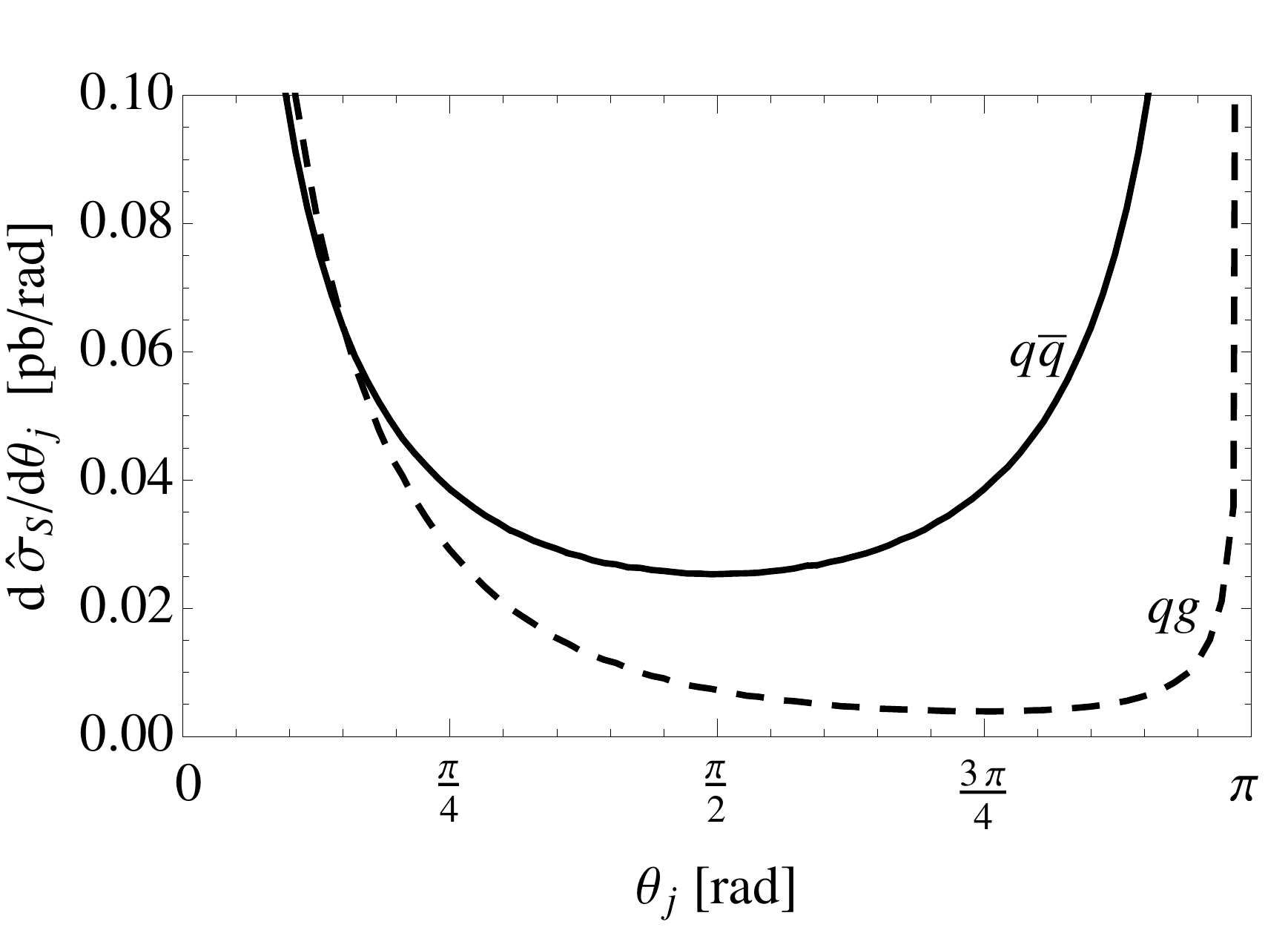}
\caption{
Symmetric part of the differential partonic cross section of 
the $q\bar{q}$ and $qg$ channels as a function of the jet 
scattering angle $\theta_{j}$ for $\sqrt{s} = 1$~TeV and 
$E_{j} \ge 20$~GeV. 
\label{fig:sigmaS_s1k}
}
\end{figure}

The above considerations are also helpful for an understanding of the 
singular behavior of $\text{d}\hat{\sigma}_{A}/\text{d}\theta_{j}$ 
and $\text{d}\hat{\sigma}_{S}/\text{d}\theta_{j}$ in the collinear 
limit where a gluon is emitted at an angle\footnote{We collectively 
  denote the scattering angle of the parton which will give rise to 
  a jet by $\theta_{j}$.} $\theta_{j} \to 0, \, \pi$. 
In SM QCD, the symmetric part of the differential cross section, 
$\text{d}\hat{\sigma}_{S}^{g}/\text{d}\theta_{j}$, shown in 
Fig.~\ref{fig:sigmaS_s1k}, is divergent for $\theta_{j} \to 0, \, \pi$, 
while the anti-symmetric cross section must be finite in this 
limit\footnote{The charge 
   asymmetry for the inclusive $t\bar{t}$ production is generated 
   at NLO. Collinear singularities appearing in the NLO calculation 
   have to be factorized and can be absorbed into PDFs only if they 
   have the same symmetry properties as the LO cross section. Since 
   the LO $q\bar{q} \to t\bar{t}$ diagrams are symmetric under charge 
   conjugation, collinear divergences can only contribute to the 
   symmetric part of the cross section \cite{Kuhn:1998kw}.
   \label{footnote:asym-at-nlo}}. 
Therefore the asymmetry $\hat{A}^{SM}(\theta_{j}) = 
(\text{d}\hat{\sigma}_{A}^{g}/\text{d}\theta_{j}) / 
(\text{d}\hat{\sigma}_{S}^{g} / \text{d}\theta_{j})$ vanishes at 
LO in $t\bar{t}g$ production for $\theta_{j} \to 0, \, \pi$ 
\cite{Berge:2013xsa}. Since a large part of the total cross 
section comes from the collinear region, the normalized asymmetry 
is suppressed, but can be enhanced considerably by appropriate 
phase space cuts.  

If axigluons are present, the different interference terms from 
$\hat{\sigma}_{A}^{gG}$ and $\hat{\sigma}_{S}^{gG}$ contribute 
in the opposite way to the symmetric and anti-symmetric parts of 
the cross section as for the SM interference terms. Therefore, 
$\text{d}\hat{\sigma}_{A}^{gG}/\text{d}\theta_{j}$ is divergent 
for $\theta_{j} \to 0, \pi$ while $\text{d}\hat{\sigma}_{S}^{gG} 
/\text{d}\theta_{j}$ is finite in this limit. As will be shown in 
the next section, the resulting asymmetry $\hat{A}^{C}(\theta_j)$ 
is not suppressed for $\theta_{j} \to 0, \pi$ if axigluons are 
present and one finds large differences of the charge asymmetry 
compared with the SM. Then also the total integrated asymmetry 
$\hat{A}^{C}_{\text{int}}$ is not suppressed in the collinear region 
and no phase space cuts are needed to search for deviations from 
the SM prediction.

\section{Parton Level Results
\label{sub:Parton-level-distributions}
}

In this section we investigate the impact of light axigluons on the 
energy, rapidity and incline asymmetries at the parton level. 
This study will help us to identify suitable cuts for observables 
at the hadron level which we will describe in the next section. In 
order to be able to compare with previous results~\cite{Berge:2013xsa}, 
we choose\footnote{The qualitative behavior of the asymmetries depends 
  only mildly on $\sqrt{\hat{s}}$.} 
$\sqrt{\hat{s}} = 1$~TeV and we always apply a cut on the energy of 
the parton (quark or gluon) leading to a jet, $E_{j} \ge 20$~GeV. 
We separate the SM prediction $\hat{\sigma}_{A}^{C,SM} = 
\hat{\sigma}_{A}^{C,g}$  from 
the anti-symmetric part of the cross sections 
and the SM prediction $\hat{A}^{C,SM}_{\text{int}}$ from the charge asymmetry.
We will display numerical results for the differences 
$\Delta\hat{\sigma}_{A}^C$ and $\Delta\hat{A}^{C}_{\text{int}}$ due to 
the presence of axigluons: 
\begin{eqnarray*}
&&
\Delta\hat{A}^{C}_{\text{int}} 
= 
\hat{A}^{C}_{\text{int}} - \hat{A}^{C,SM}_{\text{int}} 
\quad {\rm with} \quad
\hat{A}^{C}_{\text{int}}
=  
\frac
{\hat{\sigma}_{A}^{C,g} + \hat{\sigma}_{A}^{C,gG} + \hat{\sigma}_{A}^{C,G}}
{\hat{\sigma}_{S}^{g} + \hat{\sigma}_{S}^{gG} + \hat{\sigma}_{S}^{G}} 
\quad {\rm and} \quad
\hat{A}^{C,SM}_{\text{int}} 
= 
\frac{\hat{\sigma}_{A}^{C,g}}{\hat{\sigma}_{S}^{g}} 
\, ,
\\
&&
\Delta\hat{\sigma}_{A}^C 
= 
\hat{\sigma}_{A}^C - \hat{\sigma}_{A}^{C,SM} 
.
\end{eqnarray*} 
Differential cross sections and asymmetries are separated into their 
different contributions in an analogous way. To simplify the notation, 
we use from now on the short labels $C = E$, $y$ and $\varphi$ to 
indicate the energy, rapidity and incline asymmetries. The 
anti-symmetric cross sections and asymmetries are, however, defined 
as explained above (see Eq.~(\ref{eq:sigma_A_partdiff_C})) by 
imposing the conditions $\Delta E > 0$, $\Delta y > 0$ and 
$\cos\varphi > 0$, respectively. 

\begin{figure}[t]
\includegraphics[height=5.4cm]{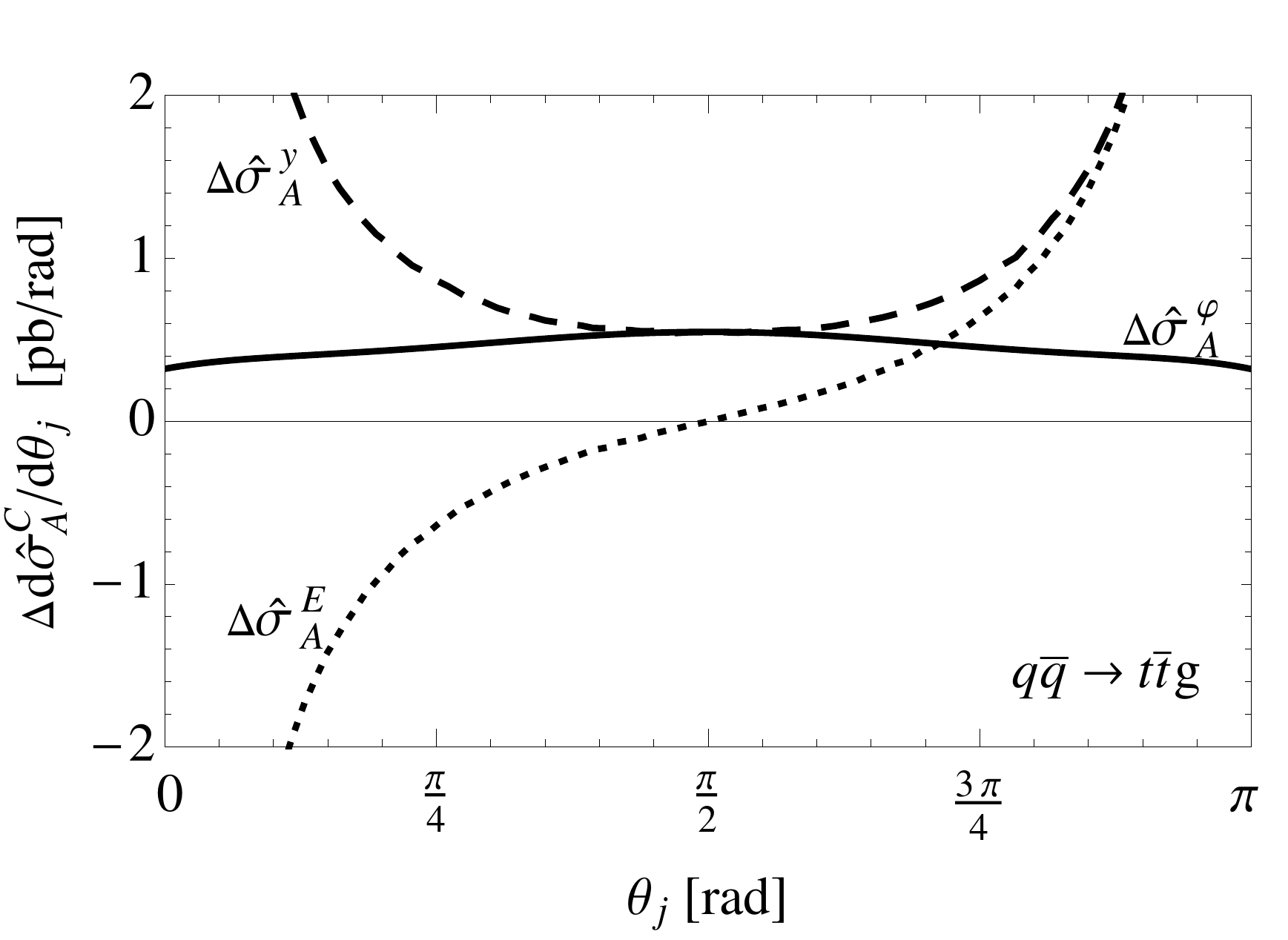}
\hspace{1mm}
\includegraphics[height=5.4cm]{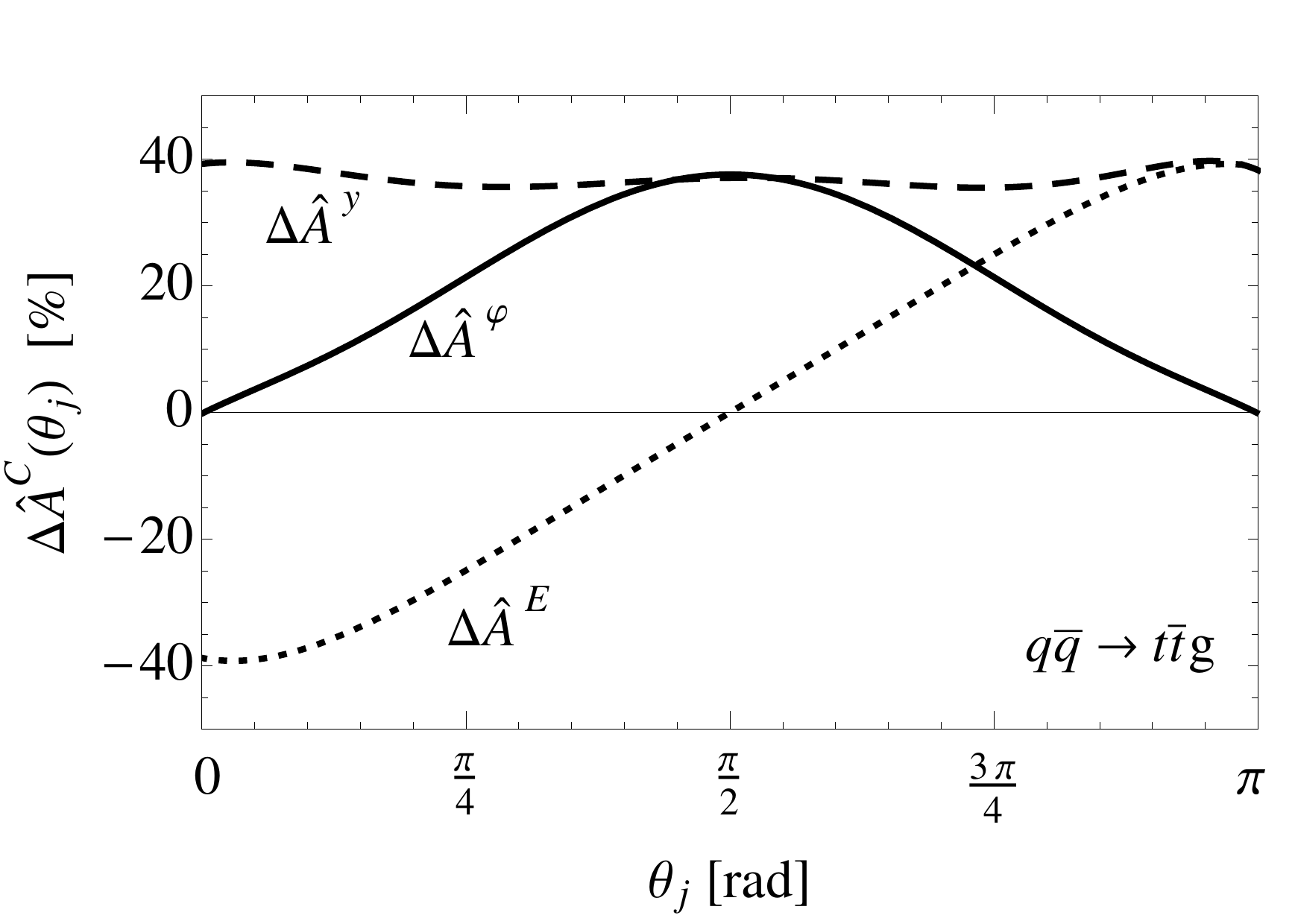}
\caption{
Partonic energy asymmetry (dotted 
lines), rapidity asymmetry (dashed lines) and incline asymmetry (solid
lines)
for the $q\bar{q}$ 
channel as a function of the jet scattering angle $\theta_{j}$ 
for $\sqrt{\hat{s}} = 1$~TeV and $E_{j} \ge 20$~GeV. Only the 
difference of the charge asymmetry with and without axigluons in 
the large-coupling scenario is shown. Left panel: $\Delta \text{d} 
\hat{\sigma}_{A}^C / \text{d}\theta_{j}$, right panel: 
$\Delta\hat{A}^C(\theta_{j})$. 
\label{fig:s1k_EA_qq}
}
\end{figure}

Figure~\ref{fig:s1k_EA_qq} shows $\Delta\text{d}\hat{\sigma}_{A}^C 
/ \text{d}\theta_{j}$ (left panel) and $\Delta\hat{A}^{C}(\theta_{j})$ 
(right panel) as a function of the jet scattering angle $\theta_{j}$ 
for the $q\bar{q} \to t\bar{t}g$ channel. The dotted lines correspond 
to the energy asymmetry, the dashed lines to the rapidity asymmetry 
and the solid lines to the incline asymmetry. Here we have chosen 
the large-coupling scenario with $m_{A} = $400~GeV, $\alpha_{A} = 
0.032$ and $\Gamma_{A} = 0.1 \cdot m_{A}$. For these parameter values, 
the interference term $\hat{\sigma}^{gG}_A$ dominates the axigluon 
contributions; therefore the dependence on $\alpha_{A}$ is linear 
to a good approximation and predictions for 
$\Delta\text{d}\hat{\sigma}_{A}^C / \text{d}\theta_{j}$ and 
$\Delta\hat{A}^{C}(\theta_{j})$ for smaller values of the axigluon 
coupling can be inferred from these figures by scaling down the 
results by an appropriate factor.  

The dotted line in Fig.~\ref{fig:s1k_EA_qq} shows the dependence
of the energy asymmetry on the jet scattering angle. The anti-symmetric 
part of the cross section $\text{d}\hat{\sigma}_{A}^{E} / 
\text{d}\theta_{j}$ exhibits a collinear divergence if axigluons are 
taken into account (see Sec.~\ref{sec:Theory}). This is an important 
and outstanding difference to the SM prediction where the 
anti-symmetric cross section vanishes for collinear jets. As in the 
case of the SM prediction, $\text{d}\hat{\sigma}_{A}^{E} / 
\text{d}\theta_{j}$ has opposite signs for $\theta_{j} < 
\frac{\pi}{2}$ and $\theta_{j} > \frac{\pi}{2}$ and is zero for 
$\theta_{j} = \frac{\pi}{2}$. In order to construct a non-vanishing 
integrated asymmetry, one therefore needs to split the integration 
region for the $q\bar{q}$-channel as it was suggested for the 
integrated SM $q\bar{q}$-asymmetry in Ref.~\cite{Berge:2013xsa}. 
The normalized energy asymmetry, shown in Fig.~\ref{fig:s1k_EA_qq} 
(right panel), has a non-zero value for collinear jet emission 
with $\Delta\hat{A}^{E}(\theta_{j})$ reaching approximately $\mp 
40\%$ for $\theta_j \to 0, \pi$. For an integrated asymmetry, one 
therefore is not forced to apply cuts on $\theta_{j}$ to suppress 
the collinear region as in the case of the SM ~\cite{Berge:2013xsa}.
It is also important to note that the shape of the energy asymmetry 
as a function of $\theta_{j}$ is quite different from the SM 
prediction.  This might be helpful in distinguishing an axigluon 
scenario from other models, provided the event rate is large enough. 

The $\theta_{j}$-dependence of the rapidity asymmetry difference
for the partonic $q\bar{q} \to t\bar{t}g$ channel is shown by the
dashed lines in Fig.~\ref{fig:s1k_EA_qq}. As in the case of the
energy asymmetry, the rapidity asymmetry exhibits a collinear 
divergence for $\theta_{j} \to 0, \pi$ which is absent 
for the SM prediction. The asymmetry difference is always positive. 
The normalized asymmetry difference $\Delta\hat{A}^{y}(\theta_{j})$ 
is therefore finite and also positive for all jet scattering angles 
and varies only very little with $\theta_{j}$. For the large-coupling 
scenario it amounts to $\Delta\hat{A}^{y}(\theta_{j}) \approx 40 \%$. 
Also here, there is no need to cut out collinear jet emission and 
the full event sample can be used for the search for axigluons.

The solid line in the left part of Fig.~\ref{fig:s1k_EA_qq} shows 
$\Delta\text{d}\hat{\sigma}_{A}^{\varphi} / \text{d}\theta_{j}$ 
for the incline asymmetry in the large-coupling scenario. The 
difference is always positive and reaches its maximum for central 
jet emission. However, as in the SM case, the dependence on the jet 
scattering angle is small. For  the asymmetry difference of the 
incline asymmetry, $\Delta\hat{A}^{\varphi}(\theta_{j})$, in 
Fig.~\ref{fig:s1k_EA_qq}, right part, one observes a suppression 
in the collinear regime. The reason is that the symmetric SM cross 
section in the denominator of the definition of $\hat{A}$ is 
divergent for $\theta_{j} \to 0, \pi$ \cite{Berge:2013xsa}. The 
asymmetry difference is largest for central jet emission and, for 
the large-coupling scenario, amounts to about 
$\Delta\hat{A}^{\varphi}(\theta_{j}) \approx 40\%$. As the SM 
prediction is roughly $\hat{A}^{\varphi,SM}(\theta_{j}) \approx 
-40\%$ for central jet emission, the total asymmetry 
$\hat{A}^{\varphi}(\theta_{j})$ for the large-coupling scenario 
would be between $0$ and $-4\%$ over the entire range of the jet 
scattering angle. For smaller coupling parameters $\alpha_{A}$ the 
total asymmetry would lie between this value and the SM prediction of 
$\hat{A}^{\varphi,SM}(\theta_{j}) \approx -40\%$ for central jets. 
Therefore, measuring a zero incline asymmetry would be indicative 
of the existence of axigluons and, provided the precision is good 
enough, be in disagreement with the SM prediction.

In Fig.~\ref{fig:s1k_EA_qq} one can see that the anti-symmetric 
parts of the cross section entering the energy and rapidity 
asymmetries are divergent for $\theta_j \to 0$, $\pi$, but the 
anti-symmetric part $\Delta \text{d}\hat{\sigma}_A^{\varphi} / 
\text{d}\theta_{j}$ for the incline asymmetry is finite in these 
limits. This absence of a collinear pole in the anti-symmetric 
$t\bar{t}+jet$ cross section is again related to the symmetry 
properties of the corresponding inclusive $t\bar{t}$ production 
cross section (see Sec.~\ref{sec:Theory} and footnote 
\ref{footnote:asym-at-nlo} at the end of Sec.~\ref{sec:Theory}) 
and a divergent behaviour in the leading order $t\bar{t}+jet$ 
cross section is connected with a collinear pole proportional 
to $1/\epsilon$ in the NLO cross section for the $t\bar{t}$ cross 
section (see, e.g., Eq.\ (3.15) in \cite{Bernreuther:2004jv}). 
Since the $2 \to 2$ Born cross section is invariant under rotations 
about the beam axis, there is no collinear pole in the 
$\varphi$-anti-symmetric part of the NLO $t\bar{t}$ cross section 
and correspondingly the incline asymmetry of the $t\bar{t}+jet$ 
final state stays finite in the collinear limits. On the other 
hand, the $q\bar{q} \to t\bar{t}$ Born cross section is 
anti-symmetric with respect to $\Delta y$ if axigluons are 
present, and collinear poles contribute to its NLO corrections. 
Correspondingly, $\text{d}\hat{\sigma}_{A}^y / \text{d}\theta_{j}$ 
is divergent for $\theta_j \to 0$, $\pi$ when axigluons are present. 
The anti-symmetry in the rapidity is also reflected in the energy 
asymmetry since the rapidities of the top and anti-top quark are  
correlated by kinematics with $\Delta E$.

\bigskip

\begin{figure}[t]
\includegraphics[height=5.4cm]{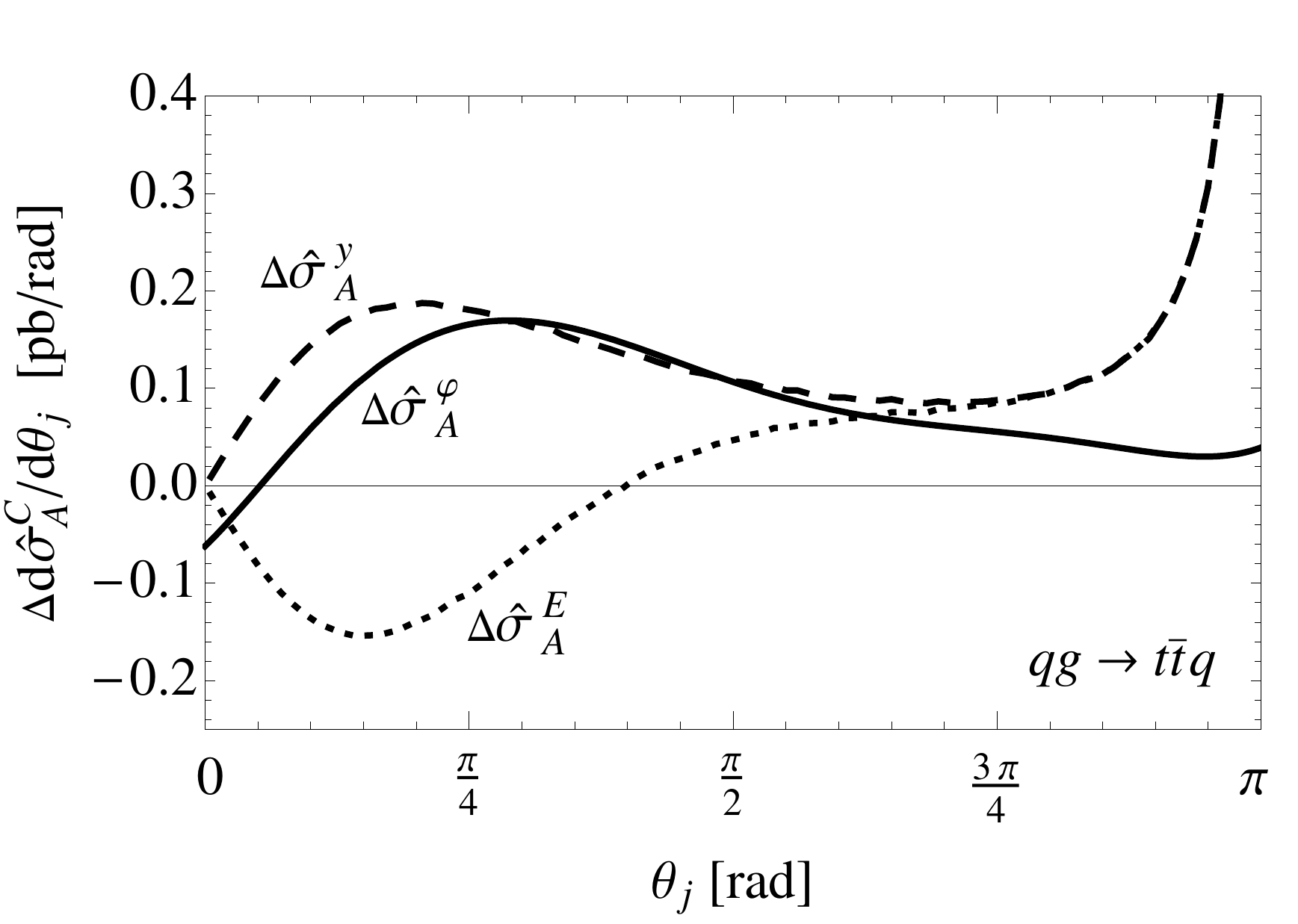}
\includegraphics[height=5.4cm]{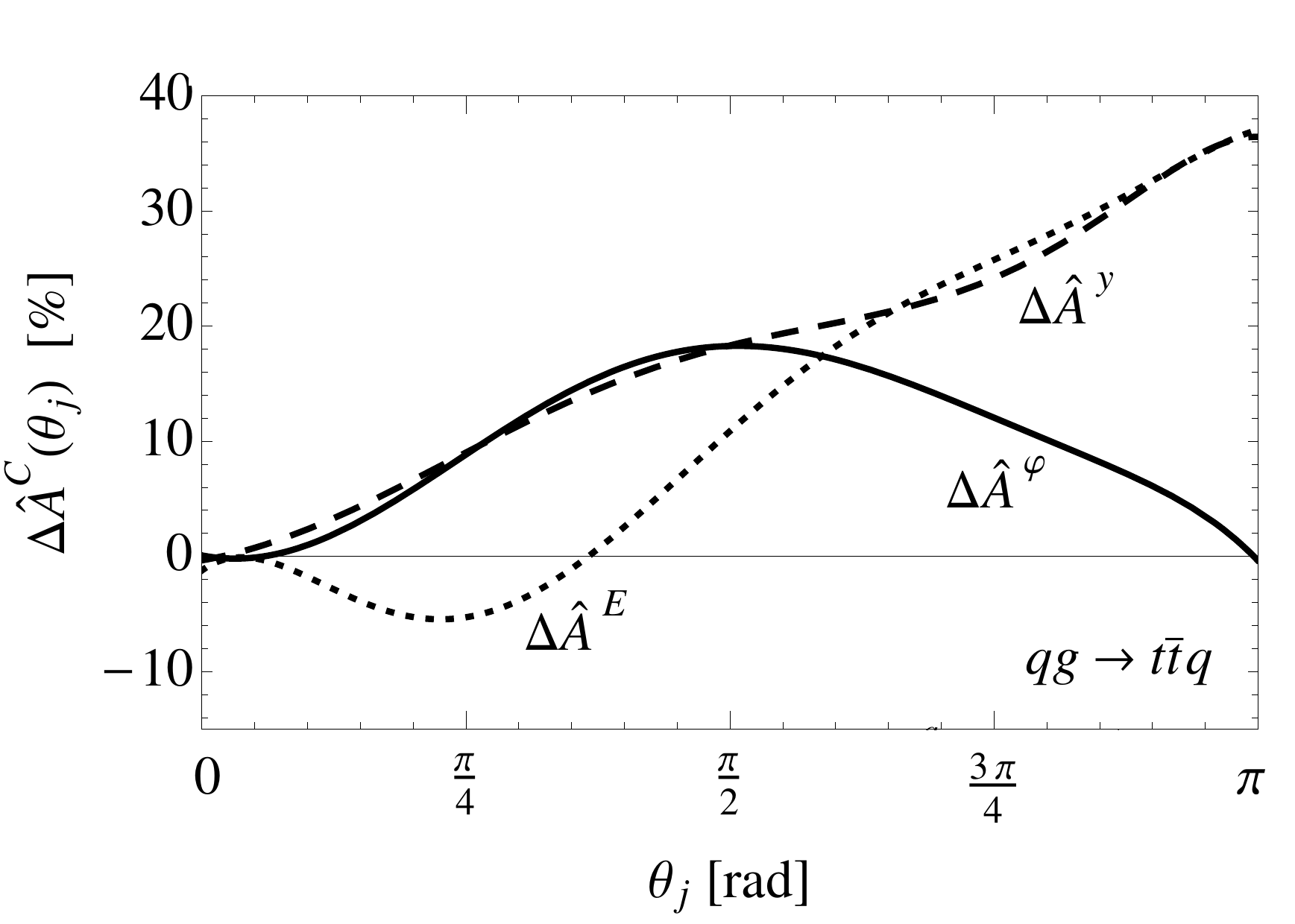}
\caption{
Partonic energy asymmetry (dotted lines), rapidity asymmetry 
(dashed lines) and incline asymmetry (solid lines) for the $qg$ 
channel as a function of the jet scattering angle $\theta_{j}$ 
for $\sqrt{\hat{s}} = 1$~TeV and $E_{j} \ge 20$~GeV. We show the 
difference of the charge asymmetry with and without axigluons for 
the large-coupling scenario. 
Left panel: $\Delta \text{d} \hat{\sigma}_{A} / \text{d}\theta_{j}$, 
right panel: $\Delta\hat{A}^C(\theta_{j})$. 
\label{fig:s1k_EI_qq_qg}
}
\end{figure}

Now we will discuss the contribution of the $qg \to t\bar{t}q$ 
channel to the various charge asymmetries. Figure~\ref{fig:s1k_EI_qq_qg} 
shows $\Delta\text{d}\hat{\sigma}_{A} / \text{d}\theta_{j}$ 
(left panel) and $\Delta\hat{A}(\theta_{j})$ (right panel) for 
this case. Again, the dotted lines denote the energy asymmetry,
the dashed lines the rapidity asymmetry and the
solid lines the incline asymmetry. 

\begin{figure}[!t]
\begin{centering}
~~~~~
\includegraphics[height=2.4cm]{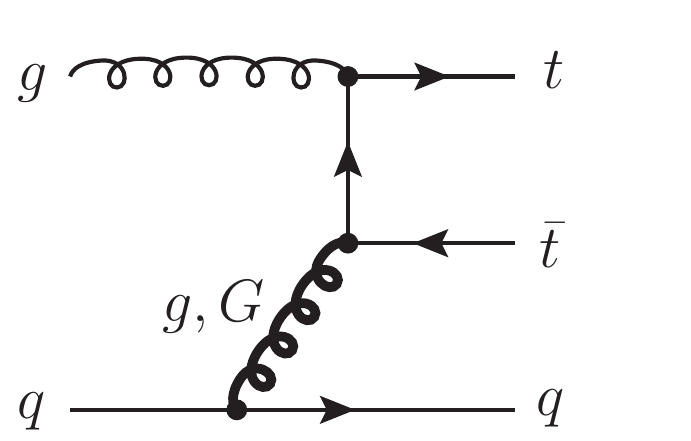}
\hspace*{.6cm}
\includegraphics[height=2.4cm]{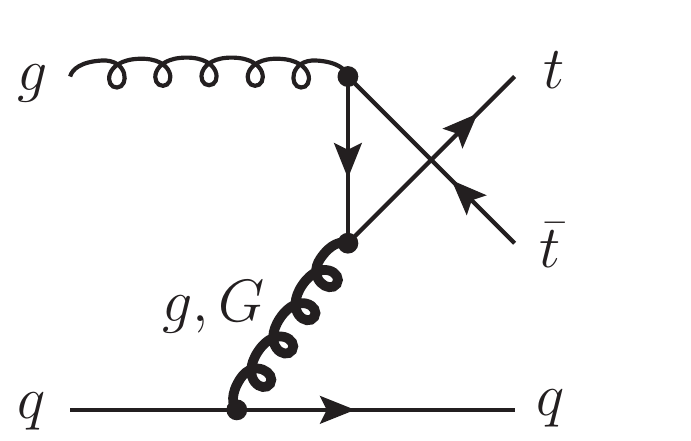}
\hspace*{.6cm}
\includegraphics[height=2.4cm]{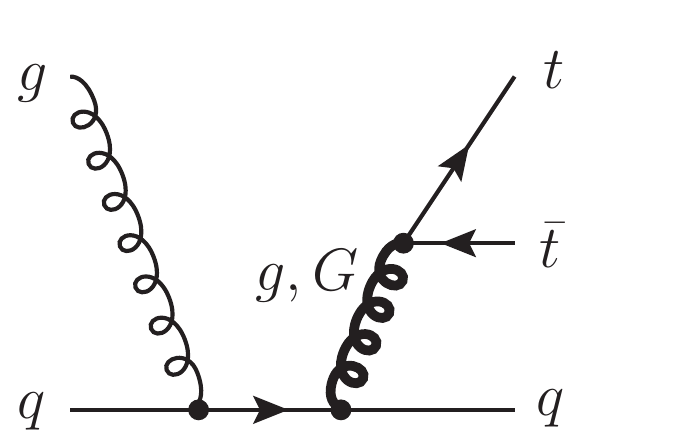}
\hspace*{.6cm}

{\footnotesize (a)}
\hspace{3.8cm}
{\footnotesize (b)}
\hspace{3.8cm}
{\footnotesize (c)}
\end{centering}

\begin{centering}
\includegraphics[height=2.4cm]{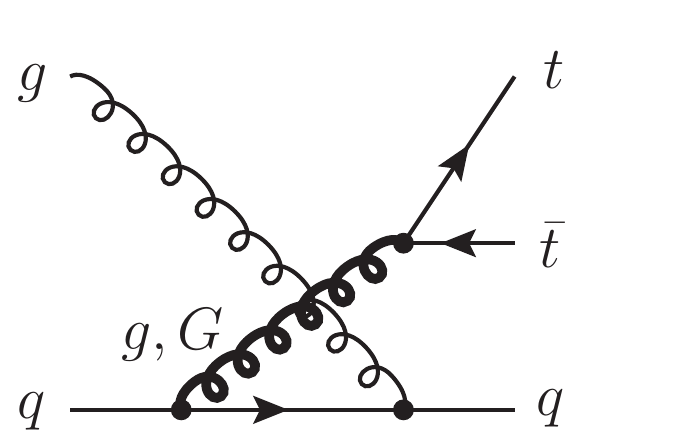}
\hspace*{.6cm}
\includegraphics[height=2.4cm]{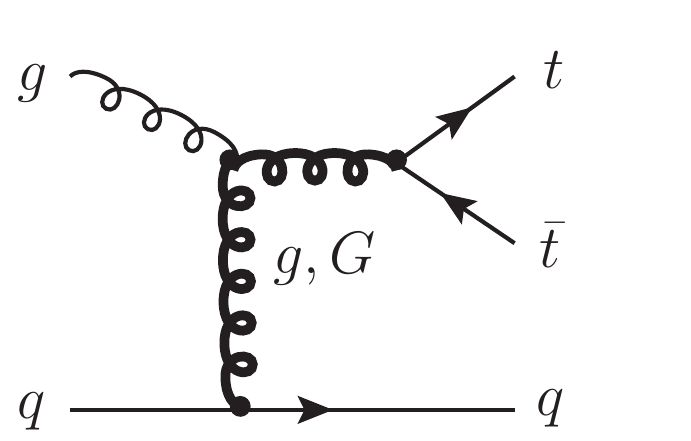}

{\footnotesize (d)}
\hspace{4.2cm}
{\footnotesize (e)}
\end{centering}
\caption{
\label{fig:qg_ttj_axi} 
Feynman diagrams contributing to the partonic process $qg
\rightarrow t\bar{t}q$. Thin curly lines denote SM gluons while
thick curly lines can represent either a SM gluon ($g$) or an 
axigluon ($G$).
}
\end{figure}

The Feynman diagrams for the $qg$ sub-process are shown in 
Fig.~\ref{fig:qg_ttj_axi}. In the SM, the symmetric part of the 
cross section for the $qg \to t\bar{t}q$ channel exhibits a 
$t$-channel singularity for $\theta_{j} \to 0$ coming from the 
Feynman diagrams in Fig.~\ref{fig:qg_ttj_axi}a, b and e. There is 
also a less strong $u$-channel singularity from the diagram in 
Fig.~\ref{fig:qg_ttj_axi}d generating a peak at $\theta_{j} \to \pi$ 
(see Fig.~\ref{fig:sigmaS_s1k}). The behaviour of the anti-symmetric 
parts of the $qg$ cross section in the collinear limits can be 
understood with similar arguments as above. The incline asymmetry 
stays finite for $\theta_{j} \to 0, \pi$ since the $2 \to 2$ Born 
cross section is invariant under rotations about the beam axis 
and collinear poles cancel in the anti-symmetric part 
$\Delta\text{d}\hat{\sigma}_{A} / \text{d}\theta_{j}$. The 
$t$-channel pole at $\theta_{j} \to 0$ cancels also in the 
rapidity and the energy asymmetry since it can be factorized 
into the $gg$-channel Born cross section which is symmetric 
with respect to $\Delta y$. In contrast, the $u$-channel pole 
survives since it is factorized into the $q\bar{q}$-channel 
Born cross section which is anti-symmetric with respect to 
$\Delta y$ if axigluons are present. 

From the difference of the energy asymmetry $\Delta\text{d} 
\hat{\sigma}_{A}^{E}(\theta_{j}) / \text{d}\theta_{j}$ shown in 
Fig.~\ref{fig:s1k_EI_qq_qg} we conclude again that the presence of 
axigluons leads to a quite different $\theta_{j}$ dependence compared 
with the SM prediction. $\Delta\text{d} \hat{\sigma}_{A}^{E}(\theta_{j}) 
/ \text{d}\theta_{j}$ is negative for $\theta_{j} \lsim 2\pi/5$ and 
positive for $\theta_{j} \gsim 2\pi/5$. Furthermore, $\Delta\text{d} 
\hat{\sigma}_{A}^{E}(\theta_{j}) / \text{d}\theta_{j}$ shows the 
same $u$-channel divergence as the symmetric part of the cross 
section for $\theta_{j} \to \pi$. However, contrary to the symmetric 
part of the SM cross section, there is no divergence for 
$\theta_{j} \to 0$. The normalized energy asymmetry difference 
$\Delta\hat{A}^{E}(\theta_{j})$ is therefore zero for $\theta_{j} 
\to 0$, but finite and large for $\theta_{j} \to \pi$ with 
$\Delta\hat{A}^{E}(\theta_{j} = \pi) \approx 35\%$. In the 
large-coupling scenario, the total asymmetry $\hat{A}^E(\theta_j)$ (not shown
in Fig.~\ref{fig:s1k_EI_qq_qg})  changes 
sign at around $\theta_j \approx \pi/2$. A non-vanishing integrated 
energy asymmetry in the $qg$-channel can therefore only be obtained 
if the integration region is separated into regions with a definite 
sign, e.g.\ with $\theta_{j} < \pi/2$ and $\theta_{j} > \pi/2$.

The SM predictions for $\text{d}\hat{\sigma}_{A}^{y,SM} / \text{d} 
\theta_{j}$ and $\hat{A}^{y,SM}(\theta_{j})$ in the $qg$-channel are 
tiny and the corresponding differences $\Delta \text{d}
\hat{\sigma}_{A}^{y}  / \text{d} \theta_{j}$ and $\Delta 
\hat{A}^{y}(\theta_{j})$ shown in Fig.~\ref{fig:s1k_EI_qq_qg} are 
therefore almost identical with the predictions in the axigluon 
model. In the large-coupling scenario, $\Delta \text{d}
\hat{\sigma}_{A}^{y} / \text{d} \theta_{j}$ is large and 
positive in the entire $\theta_{j}$ range and has a $u$-channel 
singularity for $\theta_{j} \to \pi$, but approaches zero for 
$\theta_{j} \to 0$. Therefore, the normalized rapidity asymmetry 
shown in the right panel of Fig.~\ref{fig:s1k_EI_qq_qg} is zero for 
$\theta_{j} = 0$ and increases monotonically towards larger 
$\theta_{j}$ with a maximum of $\Delta\hat{A}^{y}(\theta_{j} 
= \pi) \approx 40 \%$. The rapidity asymmetry in the $qg$-channel 
is another example where the $\theta_{j}$ dependence shows a 
behavior which is markedly different from the SM prediction. 

For the incline asymmetry, the difference 
$\Delta\text{d}\hat{\sigma}_{A}^{\varphi} / \text{d} \theta_{j}$
is positive except for very small angles. 
$\Delta\hat{A}^{\varphi}(\theta_{j})$ is maximal for central jets 
and amounts to $\Delta\hat{A}^{\varphi}(\theta_{j} \approx \pi/2) 
\approx 18 \%$. The presence of collinear divergences in the 
symmetric cross section causes the normalized asymmetry 
$\Delta\hat{A}^{\varphi}(\theta_{j})$ to vanish for $\theta_{j} 
\to 0,\, \pi$. 

We have also investigated how the asymmetries depend on other 
kinematic variables. In general, the absolute values of the 
normalized asymmetries increase for larger values of $\cos\varphi$, 
$\Delta E$, or $\Delta y$. We will not go into further detail here, 
but describe additional cuts in the next section for the analysis 
at the hadron level. 

\bigskip

We end this section with a few remarks about the influence of the 
axigluon model parameters on the charge asymmetries. A variation 
of the width $\Gamma_{A}$ in the interval $[0.1 \cdot m_{A},~0.3 
\cdot m_{A}]$ was found to lead to only very small changes in general. 
The largest differences, not more than a few percent, are found in a 
small range of values of $\theta_{j}$ for the incline asymmetry in 
the $qg$ channel. We therefore fix $\Gamma_{A} = 0.1\cdot m_{A}$ for 
all subsequent calculations. The impact of different axigluon mass 
values is much larger. The difference with respect to the SM prediction 
increases with larger axigluon masses because the resonance in the 
axigluon propagator is shifted towards the top quark pair production 
threshold. For example, the anti-symmetric part of the $q\bar{q}$ 
cross section responsible for the incline asymmetry, $\text{d} 
\hat{\sigma}_{A}^{\varphi} / \text{d} \theta_{j}$, at $\theta_{j} 
= \pi/2$ changes from $-0.12$~pb for $m_{A} = 100$~GeV to $-0.02$~pb 
for $m_{A} = 400$~GeV (with $\alpha_{A} = 0.032$). On the other hand, 
axigluons also contribute to the symmetric part of the cross section. 
This contribution is small for larger mass values in the range of 
$200 - 400$~GeV, but changes dramatically for light masses of $m_{A} 
\approx 100$~GeV, where the symmetric $q\bar{q}$ cross section, 
$\text{d}\hat{\sigma}_{S} / \text{d} \theta_{j}(\theta_j=\pi /2)$, 
is enhanced by a factor of up to $5$ for large $\alpha_A$. 
This leads to a large suppression of the normalized asymmetries for 
$m_{A} \approx 100$~GeV and therefore to large differences with 
respect to the SM predictions. We find that $\Delta 
\hat{A}^{\varphi}_{\text{int}}$ at $m_{A} = 100\,{\rm GeV}$ can be 
as large as for $m_{A} = 400\,{\rm GeV}$, while it is much smaller 
for masses in the intermediate mass range. Due to the large increase 
of the symmetric part of the cross section for axigluon masses 
$m_{A} \approx 100$~GeV, it is worth to investigate prospects to 
search for axigluons in measurements of the total $t\bar{t}+jet$ 
cross section, but we do not follow this possibility here.

\section{Hadron Level Results}\label{sec:hadron-level-results}

Based on the results of the previous section we will now describe 
predictions for charge asymmetries at the hadron level, suitable 
for experiments at the LHC. In our numerical analysis, the 
factorization scale is set equal to the top-quark mass, $\mu_{f} = 
m_{t}=173.5\,\text{GeV}$ \cite{Beringer:1900zz}. All calculations 
are performed at LO QCD, using CTEQ6L1 PDFs \cite{Pumplin:2002vw} 
and the corresponding value of the strong coupling constant, 
$\alpha_{s}^{\text{LO}}(m_{t}) = 0.1180$. While we are 
not in a position to perform a full detector simulation, we 
nevertheless apply a minimal set of ``detector cuts'' suitable for 
experiments at the LHC: for the jet's transverse momentum in the 
laboratory frame, we require $p_{T}^{j} \ge 25\,\text{GeV}$ and for 
its rapidity $|y_{j}| \le 2.5$. The jet scattering angle in the 
parton CM frame can be accessed by measuring the difference of the 
rapidities $y_{j}$ and $y_{t\bar{t}j}$ in the laboratory frame. Cuts on 
the jet scattering angle $\theta_{j}$ will then be expressed in 
terms of the partonic jet rapidity\footnote{This relation is valid 
  at leading order.}, 
\begin{eqnarray}
\hat{y}_{j} = 
\frac{1}{2} \log 
\left(\frac{1 + \cos\theta_{j}}{1 - \cos\theta_{j}} \right) 
= y_{j}-y_{t\bar{t}j} 
\, .
\end{eqnarray}

As discussed above, most suitable for a search for axigluons are 
the differential charge asymmetries as a function of the jet 
scattering angle $\theta_{j}$. Such analyses will however 
require large event rates. As long as not enough data are available 
one can try to search in measurements of integrated asymmetries. In 
the following we will discuss how precisely a particular asymmetry 
has to be defined and which cuts can be used in order to optimize 
various search scenarios.

\subsection{Energy asymmetry at the hadron level}
\label{subsec:had-energy-asymmetry}

As discussed in Sec.~\ref{sub:Parton-level-distributions}, the energy 
asymmetry $\hat{A}^E(\theta_j)$ exhibits a change of sign as a function 
of the jet scattering angle with a zero at $\theta_{j} \sim \pi/2$. 
A naive definition of an asymmetry at the hadron level would then 
reduce the asymmetry since for each positive contribution 
$\sigma_A^E$ at a given angle $\theta_{j}$ from the $q\bar q$ 
initial state there is a negative contribution at the same angle 
from the $\bar q q$ initial state. At the parton level one could 
avoid this cancellation since the direction of the incoming quark 
was known. At the hadron level it is impossible to determine 
on an event-by-event basis from which of the incoming beams the quark 
and from which the anti-quark originated; however, one can enhance 
the event sample for one or the other case by taking into account that 
the valence quark distributions of the $u$ and $d$ quarks in the proton 
are dominating the sea-quark distributions for large momentum 
fractions $x$. The ratio of the momentum fractions of the two 
partons is related to the rapidity of the $t\bar t j$-system 
by $y_{t\bar{t}j} = \ln(x_{1}/x_{2})/2$, where $x_1$ is the momentum
fraction of beam 1 moving in positive $z$ direction and $x_2$ the 
momentum fraction of beam 2. Events with positive (negative) 
$y_{t\bar{t}j}$ have a higher probability to originate from a quark 
in beam 1 (beam 2) as opposed to an anti-quark in beam 1 (beam 2). 
We therefore define a $\theta_{j}$-dependent energy asymmetry by
\begin{eqnarray}
A^{E}(\theta_{j}) 
& = & 
\frac{\text{d}\sigma_{A}^{E} / 
  \text{d} \theta_{j}(\theta_{j},\, y_{t\bar{t}j}>0) 
  + 
  \text{d}\sigma_{A}^{E} / 
  \text{d} \theta_{j}(\pi-\theta_{j},\, y_{t\bar{t}j}<0)}%
 {\text{d}\sigma_{S} / \text{d} \theta_{j}} 
\, . 
\label{eq:EA_theta_j_Hadron}
\end{eqnarray}
Similarly, an integrated asymmetry can be defined by separating jet 
scattering angles in the forward hemisphere from the backward 
hemisphere according to the sign of the rapidity of the $t\bar{t}j$ 
system:
\begin{eqnarray}
A^{E}_{\text{int}} 
& = & 
\frac{\int_{\pi/2}^{\pi}\text{d}\theta_{j}\,
\text{d}\sigma_{A}^{E}/\text{d} \theta_{j}(y_{t\bar{t}j}>0) + 
\int_{0}^{\pi/2}\text{d}\theta_{j}\, 
\text{d}\sigma_{A}^{E}/\text{d} \theta_{j}(y_{t\bar{t}j}<0)}{\sigma_{S}} 
\, . 
\label{eq:AE_integrated_hadron}
\end{eqnarray}
Different choices for the range of integration over the jet angle 
are possible, but we found that the separation into forward and 
backward hemispheres as in Eq.~(\ref{eq:AE_integrated_hadron}) 
leads to largest deviations from the SM prediction, thus to 
smallest luminosities required to observe these deviations. 

\begin{figure}[t]
\includegraphics[width=7.6cm]{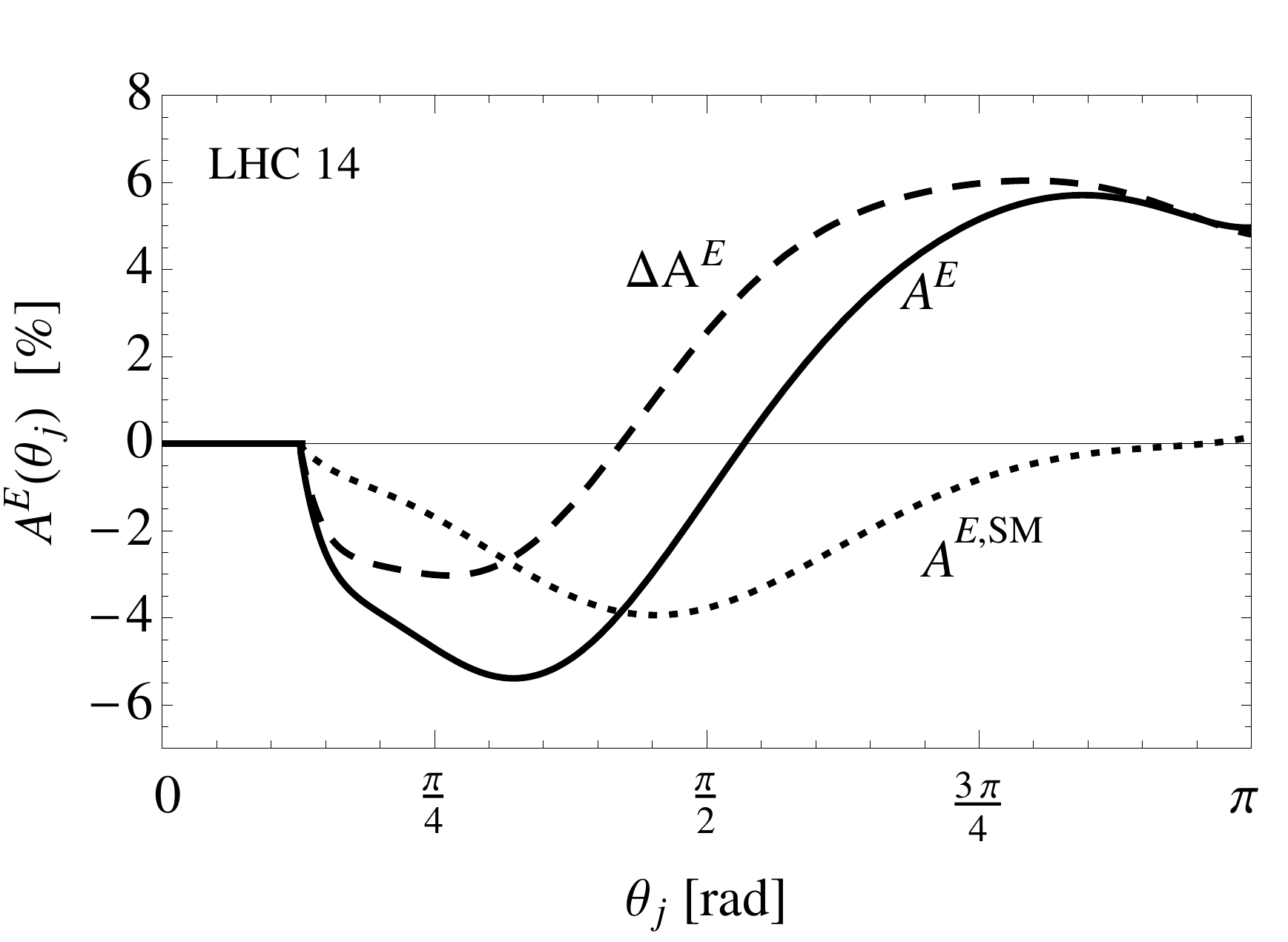}
\includegraphics[width=7.6cm]{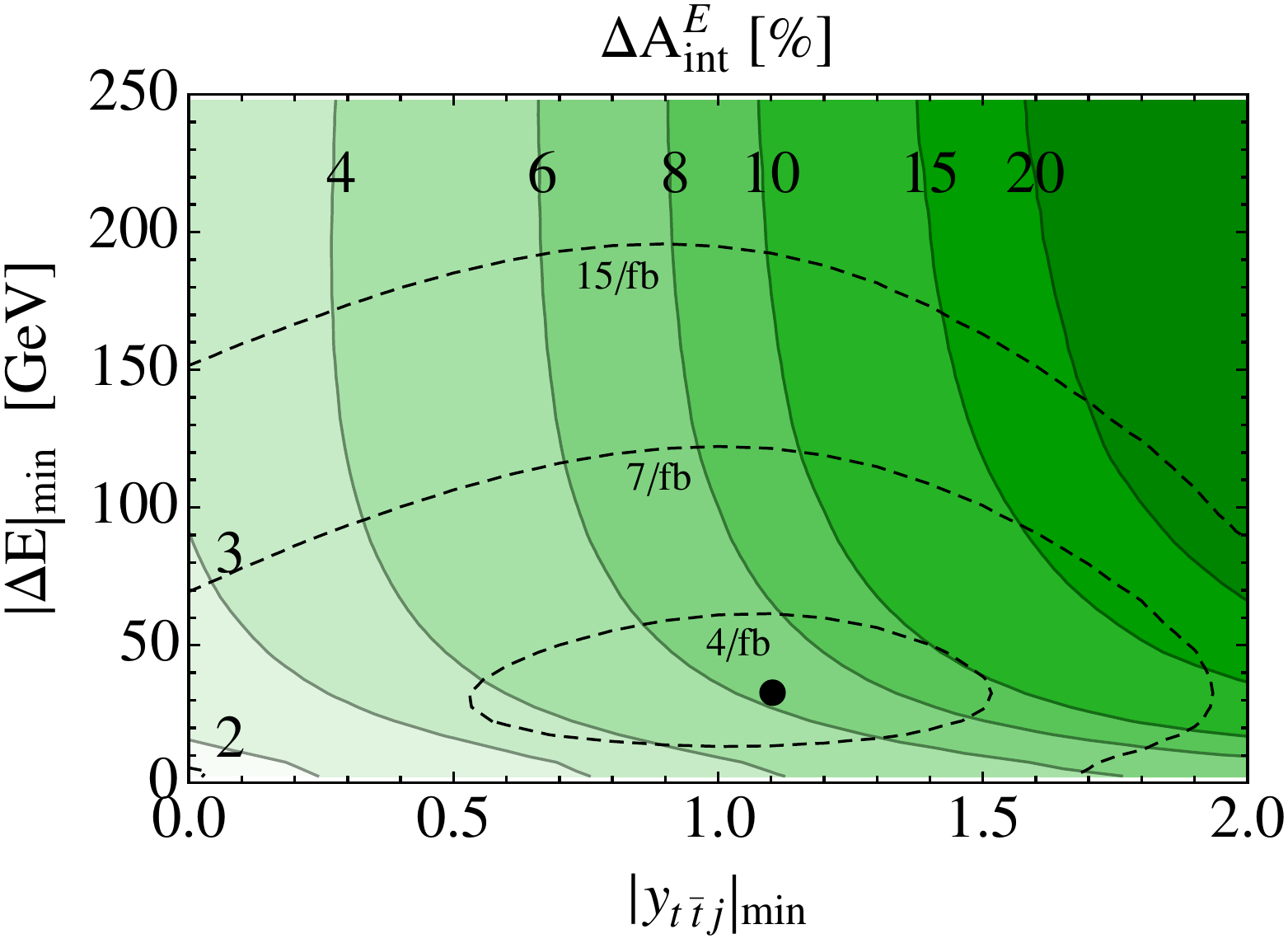}
\caption{
Energy asymmetry $A^{E}$ at the LHC with $\sqrt{S}=14$~TeV,
$m_{A}=400$~GeV, $\alpha_{A}=0.032$, $\Gamma_{A}=40$~GeV. 
Left panel: $A^{E}(\theta_j)$ (in percent) as a function of 
$\theta_j$ with $|y_{t\bar{t}j}| \geq 1$ and $|\Delta E| 
\geq 25$~GeV. The dotted line shows the SM prediction, the 
solid line includes the effect due to axigluons and the dashed 
line is the difference of both. Right panel: Contour plot of 
$\Delta A^{E}_{\text{int}}$ in percent as a function of lower 
cuts on $|y_{t\bar{t}j}|$ and $|\Delta E|$. Superimposed are dashed 
lines of constant integrated luminosity ${\cal L}$ required 
to observe $\Delta A^{E}_{\text{int}}$ at the $5\sigma$ confidence 
level. The black point indicates the optimal choice of cut values 
which lead to the minimally required luminosity.
\label{fig:LHC14_2D_energy}
}
\end{figure}

The results for $A^{E}(\theta_{j})$, $A^{E,SM}(\theta_{j})$ and 
$\Delta A^{E}(\theta_{j})$ in the large-coupling scenario are shown 
in the left panel of Fig.~\ref{fig:LHC14_2D_energy} by the solid, 
dotted and dashed lines, respectively. Since we have applied the 
``detector cut'' $|y_{j}| \le 2.5$, the asymmetries and the cross 
section are zero for $\theta_{j} \lsim \frac{\pi}{8}$. The SM 
asymmetry is always negative 
and almost identical to the results based on the definition given in 
\cite{Berge:2013xsa} which ignored the condition on the sign of 
$y_{t\bar{t}j}$ to enhance contributions related to a given 
direction of the incoming quark. The pronounced effect from the 
presence of an axigluon which was observed at the parton level 
(see Sec.~\ref{sub:Parton-level-distributions}), remains clearly 
visible also at the hadron level, as can be seen from the solid 
and dashed lines in Fig.~\ref{fig:LHC14_2D_energy} (left panel). 
For $\theta_{j} < \frac{\pi}{2}$ the asymmetry 
$A^{E}(\theta_{j})$ is negative and its absolute value 
($5.5\%$) somewhat larger than the SM prediction, however, positive 
with values up to $6\%$ for $\theta_{j} \ge \frac{\pi}{2}$. 
In this range, for $\theta_{j} \ge \frac{\pi}{2}$, the 
difference between the axigluon model and the SM is particularly 
large. Obviously, it remains true at the hadron level that the 
$\theta_{j}$ dependence of $A^{E}(\theta_{j})$ will 
be an interesting observable to search for deviations from the
SM prediction caused by the presence of axigluons. 

We have also studied the integrated asymmetry defined in 
Eq.~\eqref{eq:AE_integrated_hadron}. The SM result is negative and 
small\footnote{The SM predictions in Ref.~\cite{Berge:2013xsa} are 
  larger due to the additional upper cut $|\hat{y}_j| \le 1$ that 
  has been applied there to suppress the collinear regions.}, 
about $-0.4\,\%$ without kinematical cuts and $-2.5\%$ for the strong 
cuts $|y_{t\bar{t}j}| \geq 2$, $|\Delta E| \geq 250$~GeV. A 
contour plot for the asymmetry difference $\Delta A^{E}_{\text{int}}$ 
in the large-coupling scenario is shown in 
Fig.~\ref{fig:LHC14_2D_energy} (right panel). The difference 
$\Delta A^{E}_{\text{int}}$ is always positive and about $2\,\%$ 
without cuts on $|y_{t\bar{t}j}|$ and $|\Delta E|$, but can reach values 
of up to $20\%$ for strong cuts. $\Delta A^{E}_{\text{int}}$ has a 
strong dependence on $|\Delta E|_{min}$ for values up to 
$|\Delta E|_{min} \simeq 75$~GeV~\footnote{At LO, a lower cut on 
   $|\Delta E|$ implies a lower cut on the transverse momentum of 
   the jet. Therefore, a large value for $|\Delta E|_{min}$ is expected 
   to be advantageous also in view of a suppression of background 
   processes or higher order corrections.}. 
The dependence on $|y_{t\bar{t}j}|_{min}$ is strong for all considered 
values of $|\Delta E|_{min}$. 

The dashed lines in Fig.~\ref{fig:LHC14_2D_energy} (right panel) 
show the luminosity needed to measure the asymmetry difference 
$\Delta A^{E}_{\text{int}}$ at the $5\sigma$ level. Here we take 
into account only the statistical error $\delta A^{\varphi} = 
1/\sqrt{N}$. The number of events $N$ is calculated from the total 
hadronic cross section, $N = \sigma \cdot {\cal L} \cdot \epsilon$, 
assuming an experimental efficiency\footnote{
   In Ref.~\cite{ATLAS:2012ceu} an efficiency of $3.5\, \%$ was 
   estimated for the measurement of top-anti-top final states at 
   $\sqrt{S}=7$~TeV. We deliberately assume a slightly larger value 
   for our analysis for which there is not yet a detailed simulation 
   of the experimental conditions. In the present section we do not 
   include any background contributions. The required luminosity 
   including background can be obtained (assuming that the asymmetric 
   part of the cross section $\sigma_A$ is not altered) by rescaling 
   the efficiency $\epsilon$. Except for the case of very small 
   axi\-gluon masses, the symmetric part of the cross section $\sigma_S$ 
   is very little affected by the presence of axigluons. Then the 
   required luminosity can be approximated by ${\cal L} \approx 
   \frac{S^2\cdot \sigma_S^{SM}}{(\sigma_A-\sigma_A^{SM})^2\cdot 
   \epsilon}$\,. 
   \label{footnote-lumi}
} of $\epsilon = 5\%$. Then, using the statistical significance $S = 
|\Delta A^{\varphi}| / \delta A^{\varphi} = |\Delta A^{\varphi}| 
\sqrt{\sigma \cdot {\cal L} \cdot \epsilon}$, we determine contours 
of constant ${\cal L}$ from the condition $S=5$. We find that there 
is an optimal choice of cut values, represented by the black dot in 
this figure, where the luminosity required for a measurement at the 
$5\sigma$ level has the smallest possible value, ${\cal L}_{min} = 
3.4$~fb$^{-1}$. This value can be realized by imposing the cuts 
$|y_{t\bar{t}j}|_{min} \simeq 1.1$, $|\Delta E|_{min} \simeq 30$ GeV. A 
$5\sigma$ deviation of the energy asymmetry can be measured for the 
small-coupling scenario with ${\cal L}_{min} = 165$~fb$^{-1}$.

\subsection{Rapidity asymmetry at the hadron level}

In Sec.~\ref{sub:Parton-level-distributions} we have seen that 
the rapidity asymmetry at the parton level is not suppressed in 
the collinear regions and the whole event sample can be used. We 
therefore use the standard definition 
\begin{eqnarray}
A^{\Delta\left|y\right|}_{\text{int}} 
& = & 
\frac{\sigma_{A}^{\Delta\left|y\right|}}{\sigma_{S}} 
\quad{\rm with} \quad 
\Delta\left|y\right| = \left|y_{t}\right| - \left|y_{\bar{t}} \right| 
\label{eq:Def_Ay_hadron}
\end{eqnarray}
for the integrated rapidity asymmetry at the hadron level. It 
is defined in terms of the absolute values of the top and 
anti-top rapidities. With this definition, an experimental 
determination of the quark direction based on the boost of the 
$t \bar t j$-system is avoided. 
 
We have also seen in 
Sec.~\ref{sub:Parton-level-distributions} that at the parton 
level $\Delta\hat{A}^{y}(\theta_{j})$ has only a weak dependence 
on the jet scattering angle $\theta_{j}$ for the $q\bar{q}$ channel, 
while it increases monotonically with $\theta_{j}$ for the $qg$ 
channel. Therefore, with the definition  
\begin{eqnarray}
A^{\Delta\left|y\right|}(\theta_{j}) 
& = & 
\frac{
  \text{d}\sigma_{A}^{\Delta\left|y\right|} 
  / \text{d}\theta_{j}(\theta_{j},\, y_{t\bar{t}j}>0) 
  + 
  \text{d}\sigma_{A}^{\Delta\left|y\right|} 
  / \text{d}\theta_{j}(\pi-\theta_{j},\, y_{t\bar{t}j}<0)
  }{\text{d}\sigma_{S} / \text{d}\theta_{j}} 
\, ,
\end{eqnarray}
one can enhance the contribution from the $qg$-channel. 

\begin{figure}[t]
\includegraphics[width=7.6cm]{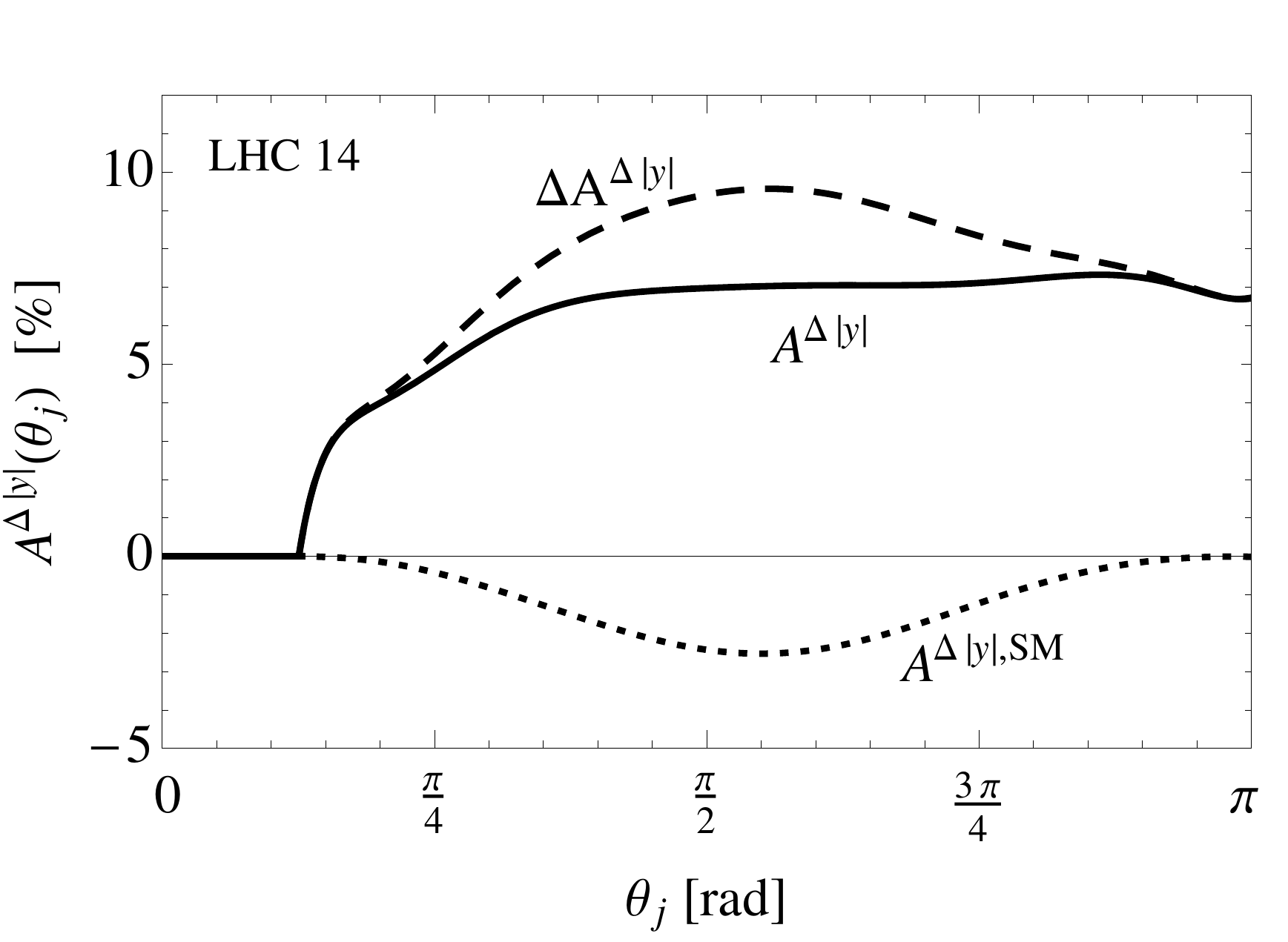}
\includegraphics[width=7.6cm]{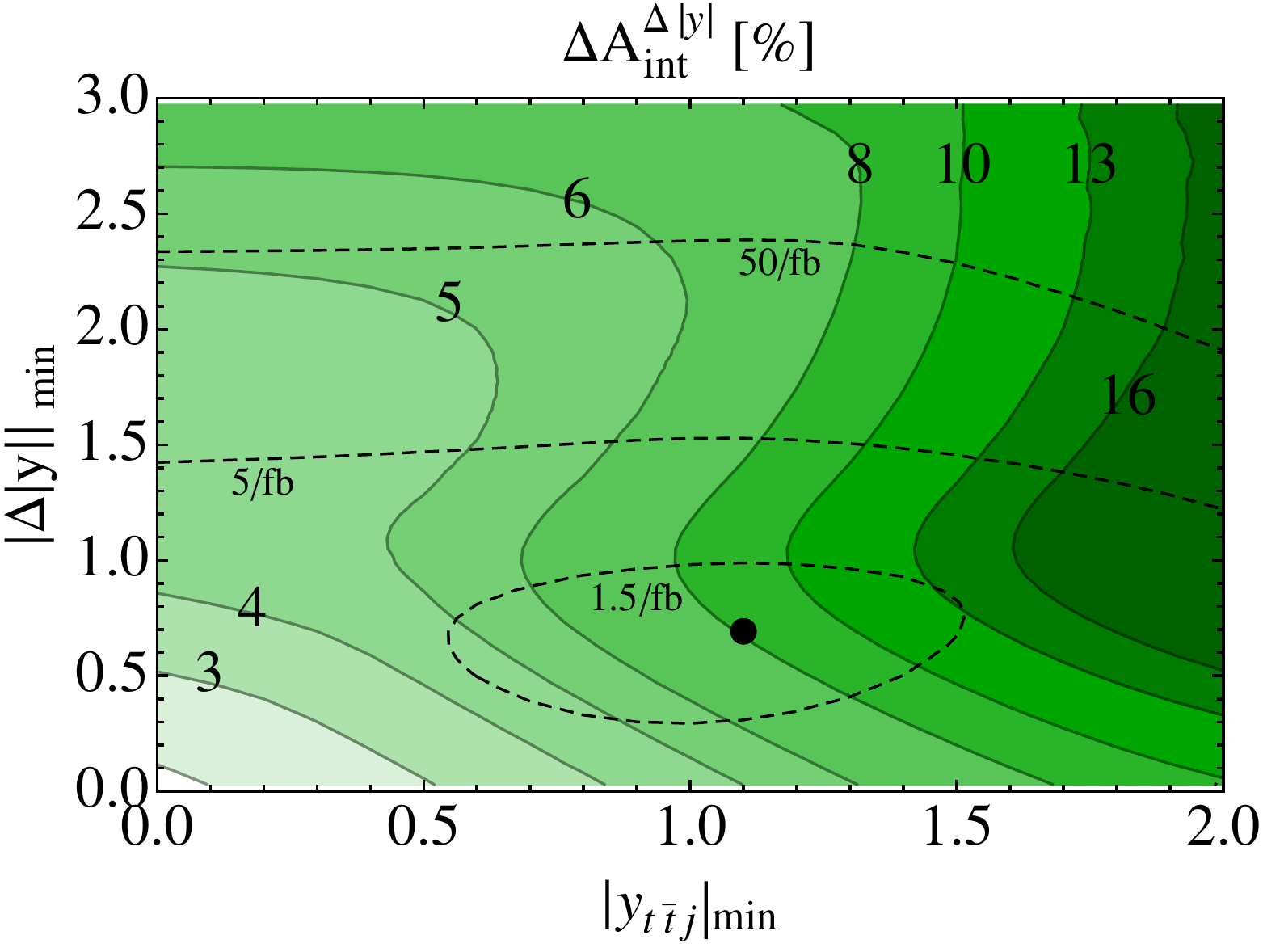}
\caption{
Rapidity asymmetry $A^{\Delta |y|}$ at the LHC with $\sqrt{S} = 
14$~TeV, $m_{A} = 400$~GeV, $\alpha_{A} = 0.032$, $\Gamma_{A} = 
40$~GeV. Left panel: $A^{\Delta |y|}(\theta_j)$ (in percent) as a 
function of $\theta_j$ with $|y_{t\bar{t}j}| > 1$ and $|\Delta\left|y\right|| 
> 0.75$. The dotted line shows the SM prediction, the solid line 
includes the effect due to axigluons and the dashed line is the 
difference of both. Right panel: Contour plot of $\Delta A^{\Delta 
|y|}_{\text{int}}$ in percent as a function of lower cuts on 
$|y_{t\bar{t}j}|$ and $|\Delta |y||$. Superimposed are dashed lines of 
constant integrated luminosity ${\cal L}$ required to observe 
$\Delta A^{\Delta |y|}_{\text{int}}$ at the $5\sigma$ confidence 
level. The black point indicates the optimal choice of cut values 
which lead to the minimally required luminosity.
\label{fig:LHC14_2D_rapidity}
}
\end{figure}

The resulting distributions for 
$A^{\Delta\left|y\right|}(\theta_{j})$,
$A^{\Delta\left|y\right|,SM}(\theta_{j})$ and 
$\Delta A^{\Delta\left|y\right|}(\theta_{j})$ are show in 
Fig.~\ref{fig:LHC14_2D_rapidity} (left panel) by the solid, 
dotted and dashed lines. The additional cuts $|y_{t\bar{t}j}| > 1$ and 
$|\Delta\left|y\right|| > 0.75$ have been applied here. The SM 
prediction is negative for all values of $\theta_{j}$ with a minimum 
of $-2.5\,\%$ at $\theta_{j} \simeq \frac{5}{9}\pi$ and tends to 
zero in the collinear region because the symmetric part of the cross 
section, i.e.\ the denominator in the definition of the rapidity 
asymmetry, increases for $\theta_{j} \to 0, \pi$. The cross section 
is zero for $\theta_{j} \lsim \frac{1}{8}\pi$ since we have applied 
the laboratory jet cut $|y_{j}| \le 2.5$. The total asymmetry
$A^{\Delta\left|y\right|}(\theta_{j})$ is large and positive
for $\theta_{j} \gsim \frac{1}{8}\pi$. The asymmetry difference 
$\Delta A^{\Delta\left|y\right|}(\theta_{j})$ is positive for 
all couplings $\alpha_{A}$. The most striking difference of the 
axigluon prediction with respect to the SM result is a non-vanishing 
asymmetry for collinear jets with $\theta_{j} \to \pi$. As before, 
the results shown in Fig.~\ref{fig:LHC14_2D_rapidity} have been 
obtained for the large-coupling scenario; for smaller couplings the 
results will scale down towards the SM prediction. 

In the right panel of Fig.~\ref{fig:LHC14_2D_rapidity} we display 
a contour plot of the integrated rapidity asymmetry in the 
large-coupling scenario as a function of lower cuts on $|y_{t\bar{t}j}|$ 
and the rapidity difference $|\Delta\left|y\right||$. The solid contour 
lines represent constant values of 
$\Delta A^{\Delta\left|y\right|}_{\text{int}}$ (in percent). If no 
additional cuts are applied, 
$\Delta A^{\Delta\left|y\right|}_{\text{int}} = 1.8\%$. This value 
can be enhanced by applying cuts on $|y_{t\bar{t}j}|$ and 
$|\Delta\left|y\right||$ to values up to 
$\Delta A^{\Delta\left|y\right|}_{\text{int}} \approx 20\%$. 
The SM result is negative and small, about $-0.3\%$ without 
kinematical cuts and $-4.8\%$ for very strong kinematical cuts of 
$|y_{t\bar{t}j}|_{\rm{min}} = 2$, $|\Delta|y||_{\rm{min}} = 3$. 
The dashed lines show again the luminosity needed to measure this 
asymmetry difference at $5\sigma$ and the black dot indicates the 
smallest possible value, ${\cal L}_{min} = 1.3$~fb$^{-1}$. A 
$5\sigma$ deviation of the rapidity asymmetry can be measured for 
the small-coupling scenario with ${\cal L}_{min} = 65$~fb$^{-1}$.

\subsection{Incline asymmetry at the hadron level}
\label{subsec:had-incline-asymmetry}

In Sec.~\ref{sub:Parton-level-distributions} we have shown that 
the incline asymmetry $\hat{A}^{\varphi}(\theta_j)$ is most strongly 
affected by axigluons, in both the $q\bar{q}$ and the $qg$ channel, 
if the jet is emitted perpendicular to the beam axis. In addition, 
$\Delta\hat{A}^{\varphi}(\theta_j)$ and $\hat{A}^{\varphi,SM} 
(\theta_j)$ is zero for collinear jets, i.e.\ in the region where 
the symmetric part of the cross section gets large. Therefore, 
contrary to the cases of the energy and rapidity asymmetries, an 
upper cut on $|\hat{y}_{j}|$ is applied to enhance the integrated 
asymmetry. Furthermore, the sign of the incline asymmetry depends 
on the direction of the incoming quark. As above, we count events 
at the reflected angle $\pi-\theta_{j}$ instead of $\theta_{j}$, 
depending on the sign of the rapidity of the $t \bar t j$-system, 
i.e., we define the integrated asymmetry by 
\begin{eqnarray}
A^{\varphi}_{\text{int}} 
& = & 
\frac{\sigma_{A}^{\varphi}(y_{t\bar{t}j}>0) - 
\sigma_{A}^{\varphi}(y_{t\bar{t}j}<0)}{\sigma_{S}} 
\label{eq:Def_Aphi_hadron}
\end{eqnarray}
and the $\theta_{j}$ dependent asymmetry
\begin{eqnarray}
A^{\varphi}(\theta_{j}) 
& = & 
\frac{
  \text{d}\sigma_{A}^{\varphi} 
  /\text{d}\theta_{j}(\theta_{j},\, y_{t\bar{t}j}>0)  
  - 
  \text{d}\sigma_{A}^{\varphi} 
  /\text{d}\theta_{j}(\pi-\theta_{j},\, y_{t\bar{t}j}<0) 
  }
  {\text{d}\sigma_{S} / \text{d}\theta_{j}}
\, .
\label{eq:IA_theta_j_Hadron}
\end{eqnarray}

This definition implies that contributions from the phase space 
region around $y_{t\bar{t}j} \approx 0$ cancel in the numerator of 
Eq.~\eqref{eq:Def_Aphi_hadron}, but fully contribute to the 
denominator. Therefore a minimum cut on the boost of the 
$t\bar{t}j$-system can strongly enhance the normalized asymmetry. 

\begin{figure}[t]
\includegraphics[width=7.6cm]{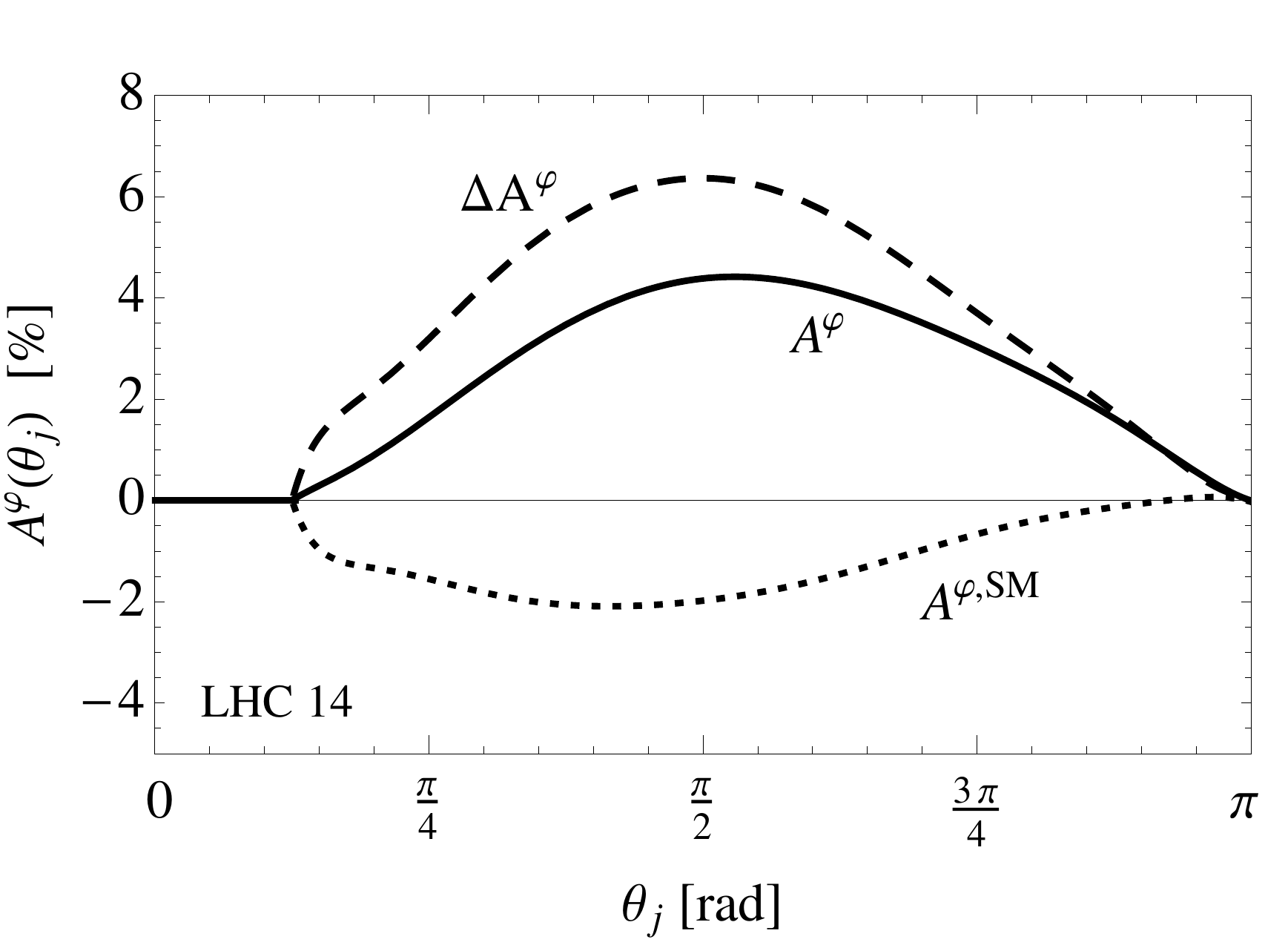}
\includegraphics[width=7.6cm]{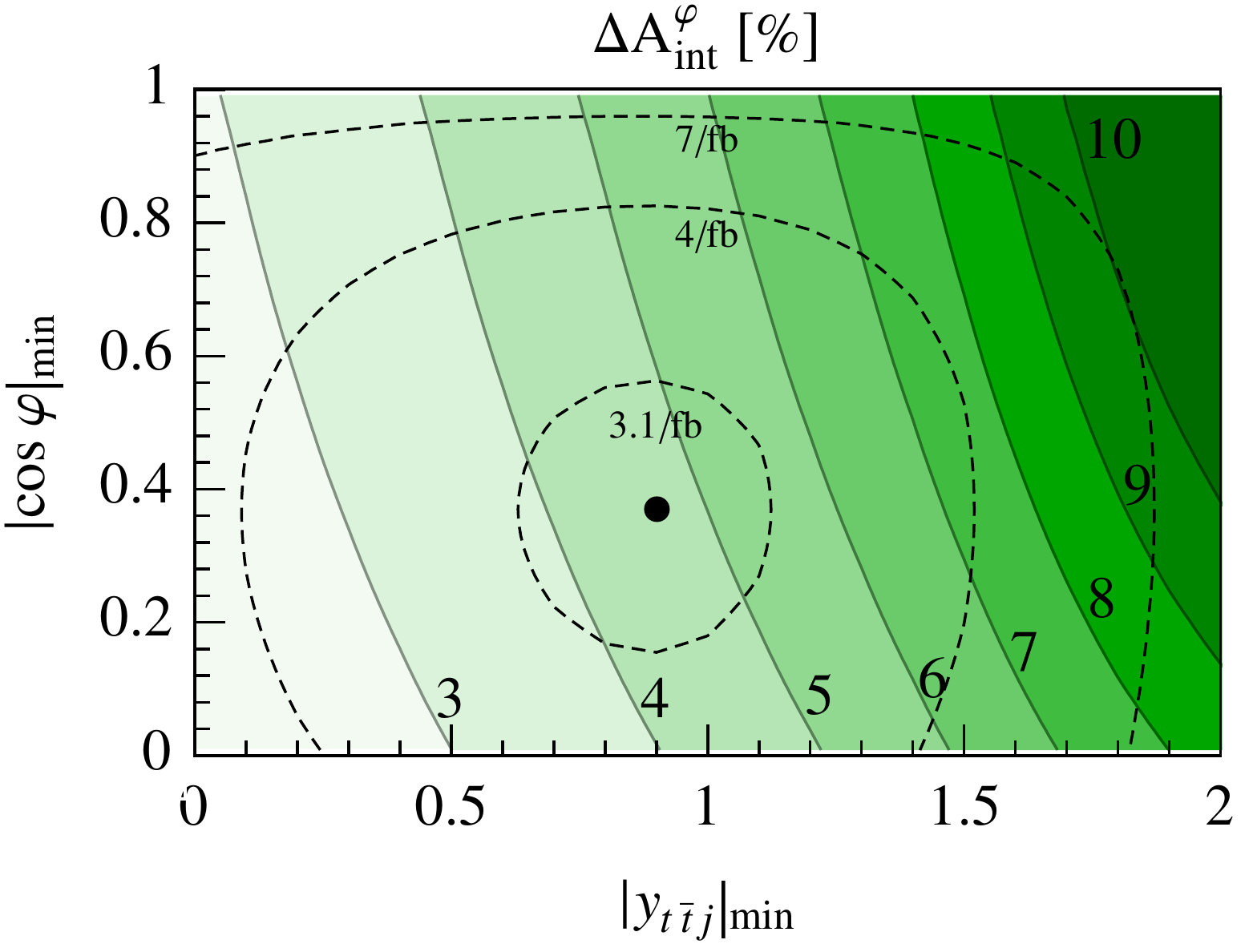}
\caption{ 
Incline asymmetry $A^{\varphi}$ at the LHC with $\sqrt{S}=14$~TeV,
$m_{A}=400$~GeV, $\alpha_{A}=0.032$, $\Gamma_{A}=40$~GeV. 
Left panel: $A^{\varphi}(\theta_j)$ (in percent) as a function of 
$\theta_j$ with $|y_{t\bar{t}j}| \geq 1$ and $|\cos \varphi| \geq 0.4$. 
The dotted line shows the SM prediction, the solid line includes 
the effect due to axigluons and the dashed line is the difference of 
both. Right panel: Contour plot of $\Delta A^{\varphi}_{\text{int}}$ 
in percent as a function of lower cuts on $|y_{t\bar{t}j}|$ and 
$|\cos\varphi|$ with $|\hat{y}_{j}| \le 1$. Superimposed are dashed 
lines of constant integrated luminosity ${\cal L}$ required to 
observe $\Delta A^{\varphi}_{\text{int}}$ at the $5\sigma$ 
confidence level. The black point indicates the optimal choice of 
cut values which lead to the minimal required luminosity.
\label{fig:LHC14_2D_incline}
}
\end{figure}

In the SM the incline asymmetry is dominated by the $q\bar{q}$ 
channel and the contribution from the $qg$ channel is small 
\cite{Berge:2013xsa}. If axigluons are present, both channels give 
positive contributions of similar size. In the left part of 
Fig.~\ref{fig:LHC14_2D_incline} we show the resulting hadronic 
asymmetry difference $\Delta A^{\varphi}(\theta_j)$ for the LHC 
with $\sqrt{S} = 14$~TeV in the large-coupling scenario. 
Fig.~\ref{fig:LHC14_2D_incline} (right panel)  
shows the  contour plot 
of $\Delta A^{\varphi}_{\text{int}}$ as a function of lower cuts on 
$|y_{t\bar{t}j}|$
and $|\cos\varphi|$. Additionally, 
an upper cut of $|\hat{y}_{j}| \le 1$ has been applied. The SM result 
is negative and small, about $-0.6\%$ without kinematical cuts and 
$-3.4\%$ for very strong kinematical cuts of $|y_{t\bar{t}j}|_{\rm{min}} 
= 2$, $|\cos\varphi|_{\rm{min}} \to 1$. The solid black lines in 
Fig.~\ref{fig:LHC14_2D_incline} (right panel) show contours of constant 
$\Delta A^{\varphi}_{\text{int}}$. We see that without cuts on 
$|y_{t\bar{t}j}|$ and $|\cos\varphi|$, asymmetry differences of 
$\Delta A^{\varphi}_{\text{int}} \approx 2\,\%$ are obtained 
while very strong cuts can bring this difference up to values 
of $\Delta A^{\varphi}_{\text{int}} \approx 10\,\%$. The 
dependence on $|\cos\varphi|_{min}$ is moderate, but a lower 
cut on $|y_{t\bar{t}j}|$ can lead to a strong enhancement. 

The dashed lines in Fig.~\ref{fig:LHC14_2D_incline} (right panel) 
show which luminosity ${\cal L}_{min}$ would be required to measure 
the predicted difference at the $5\sigma$ confidence level. For 
the special choice of parameters considered here, we find a minimal 
luminosity of ${\cal L}_{min} = 2.7$~fb$^{-1}$ where the corresponding 
asymmetry difference is $\Delta A^{\varphi}_{\text{int}} = 4.8\%$ 
for the cuts $|\cos\varphi|_{min} = 0.38$ and $|y_{t\bar{t}j}|_{min} 
= 0.95$. In the small-coupling scenario, we find 
$\Delta A^{\varphi}_{\text{int}} = 0.3 \, \%$ without cuts on 
$|\cos\varphi|$ and $|y_{t\bar{t}j}|$, and $\Delta 
A^{\varphi}_{\text{int}} \approx 2\, \%$ for strong 
$|\cos\varphi|_{min}$ and $|y_{t\bar{t}j}|_{min}$ cuts. The minimum 
luminosity required for the small-coupling scenario is ${\cal L}_{min} 
= 110$~fb$^{-1}$.

\subsection{Luminosity requirement for the measurement of charge 
asymmetries}

\begin{figure}[t]
\includegraphics[width=12cm]{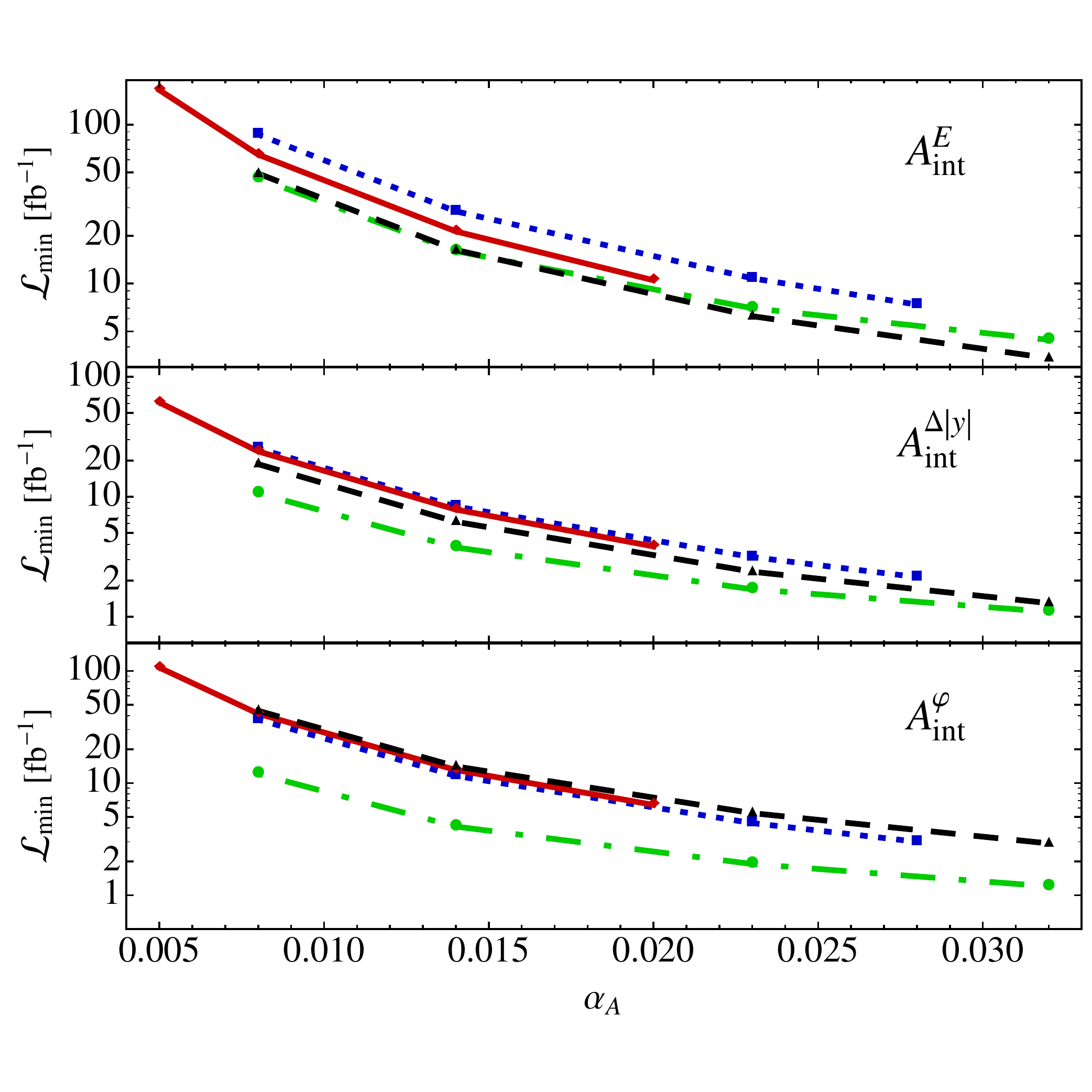}
\caption{
Minimal luminosity required to observe axigluons using the energy 
asymmetry, defined in Eq.~(\ref{eq:AE_integrated_hadron}), the 
rapidity asymmetry, Eq.~(\ref{eq:Def_Ay_hadron}), and the incline 
asymmetry, Eq.~(\ref{eq:Def_Aphi_hadron}), at the $5\sigma$ level. 
We show the dependence on the coupling parameter $\alpha_A$ at the 
abscissa. Different values of the axigluon mass $m_A={100,\, 200,\, 
300,\, 400}$~GeV correspond to the green long-dashed, blue dotted, 
red solid, and black dashed lines. The axigluon width is set to 
$0.1\cdot m_A$.
\label{fig:Lumi_LHC14}
}
\end{figure}

The analysis described in the previous sections can be performed 
for each pair of coupling strength $\alpha_A$ and mass $m_A$ and 
an optimal choice of cut values can be determined. The minimum 
required luminosity to measure the asymmetry difference 
$\Delta A_{\text{int}}$ at the $5\sigma$ level found in this way 
is presented in Fig.~\ref{fig:Lumi_LHC14} as a function of the 
coupling strength $\alpha_A$ and the considered mass values. The 
upper panel shows the results for the energy asymmetry, the 
middle panel for the rapidity asymmetry and the bottom panel for 
the incline asymmetry. The long-dashed (green) lines correspond 
to an axigluon mass of $m_A = 100$~GeV, the dotted (blue) lines 
are for $m_A =200$~GeV, the solid (red) lines for $m_A =300$~GeV 
and the dashed (black) lines for $m_A =400$~GeV. The axigluon 
width is set to $0.1\cdot m_A$. In each case we have chosen the 
range of values for the coupling strength $\alpha_A$ in accordance 
with Ref.~\cite{Gross:2012bz}. As expected, smaller luminosities 
are required for larger coupling strengths. If one restricts the 
coupling to $\alpha_{A} \ge 0.008$, then for all axigluon masses, 
the minimum required luminosities are ${\cal L}_{min} = 44$~fb$^{-1}$ 
for the incline asymmetry, ${\cal L}_{min} = 86$~fb$^{-1}$ for the 
energy asymmetry and ${\cal L}_{min} = 26$~fb$^{-1}$ for the 
rapidity asymmetry. 

In general, the integrated rapidity asymmetry is the most promising 
observable for which one obtains the smallest required luminosities. 
However, if the number of events is large enough, the energy 
asymmetry differential with respect to $\theta_j$ shows the most 
characteristic differences compared with the SM prediction as has 
been shown in Fig.~\ref{fig:LHC14_2D_energy} (left panel) above. 
A measurement of this quantity should therefore be used to search 
for axigluon contributions.

\section{LHC at $8$~TeV}\label{Sec:LHC8}

We have investigated the possibility to search for axigluons in 
the mass range of $100 - 400$~GeV using the LHC data from 2011 
at $\sqrt{S} = 7$~TeV and from 2012 at $\sqrt{S} = 8$~TeV. The 
general features of cross sections and asymmetries do not depend 
strongly on the CM energy. The most important difference at these 
lower energies is that the contribution of the gluon-gluon initial
state is smaller. Therefore the asymmetries are slightly larger and 
one may expect that already with the available data it should be 
possible to exclude large areas in the parameter space of the 
axigluon mass and coupling. 

\begin{figure}[t]
\includegraphics[width=8.5cm]{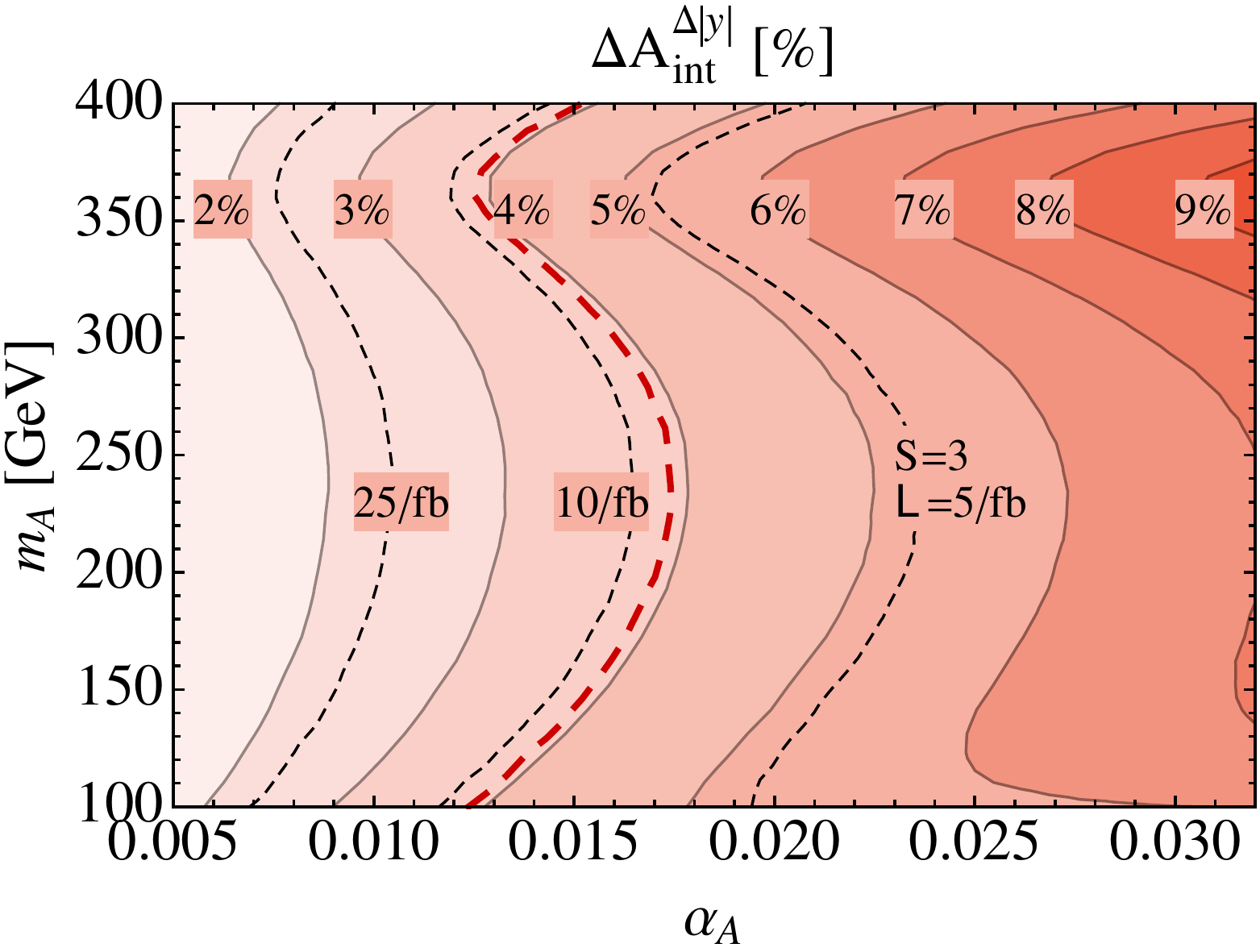}
\caption{
Difference of the rapidity asymmetry, 
$\Delta A^{\Delta|y|}_{\text{int}}$, at the LHC with $\sqrt{S} 
= 8$~TeV as a function of the coupling $\alpha_{A}$ and the 
axigluon mass $m_{A}$. The lower cuts  $|\Delta|y|| > 0.35$ and 
$|y_{t\bar{t}j}| > 0.7$ have been applied. Solid grey lines show constant 
values of $\Delta A^{\Delta|y|}_{\text{int}}$. The superimposed 
black dashed lines show which integrated luminosity ${\cal L}$ is 
needed to exclude axigluons at the $3\sigma$ confidence level. 
In the area to the right of the dashed red line, axigluons can 
be discovered in a measurement of $\Delta A^{\Delta|y|}_{\text{int}}$ 
with ${\cal L} = 25$~fb$^{-1}$ at the $5\sigma$ confidence level. 
\label{fig:LHC8_exclusion_plot}
}
\end{figure}

In the last section we have identified the integrated rapidity 
asymmetry $A^{\Delta|y|}$ as a most promising observable. We find 
that moderate cuts $|\Delta|y|| > 0.35$ and $|y_{t\bar{t}j}| > 0.7$ 
lead to the smallest integrated luminosity needed for an axigluon 
search based on $\Delta A^{\Delta|y|}_{\text{int}}$ in the mass 
range of $100 - 300$~GeV. In Fig.~\ref{fig:LHC8_exclusion_plot} we 
show $\Delta A^{\Delta|y|}_{\text{int}}$ as a function of the 
axigluon coupling $\alpha_{A}$ and mass $m_{A}$. For definiteness, 
we have chosen $\sqrt{S} = 8$~TeV and the axigluon width was fixed 
at $\Gamma_{A} = 0.1\cdot m_{A}$. The solid grey lines show constant 
values of $\Delta A^{\Delta|y|}_{\text{int}}$. Asymmetries of about 
$1.5\,\%$ for very small couplings $\alpha_{A} = 0.005$ are found 
and values of up to $\Delta A^{\Delta|y|}_{\text{int}} = 9\,\%$ can 
be reached for large couplings $\alpha_{A} = 0.032$. 

If the data for $A^{\Delta|y|}_{\text{int}}$ agree with the SM 
prediction, one can determine exclusion limits on the model 
parameters. For a rough estimate we calculate the minimal integrated 
luminosity required for exclusion limits at the $3\sigma$ level. 
Without a detailed experimental study we can base such a calculation 
only on the statistical uncertainty. We use the same cuts on 
$|\Delta|y||$ and $|y_{t\bar{t}j}|$ as given above and assume that 
$t \bar t j$ events can be observed with an efficiency of $\epsilon 
= 0.035$ \cite{ATLAS:2012ceu}. Moreover, we assume that background 
processes may increase the observed cross section by $40\, \%$ 
without affecting the asymmetric part of the cross section 
$\sigma_A$\footnote{
  See footnote \ref{footnote-lumi} in  
  Sec.~\ref{subsec:had-energy-asymmetry}.
}.
The result is shown in Fig.~\ref{fig:LHC8_exclusion_plot}. From 
the superimposed black dashed lines of constant integrated luminosity 
one can read off exclusion limits on the axigluon mass and coupling. 
We conclude that a large part of the parameter range of the light 
axigluon model can be excluded with $25$~fb$^{-1}$ of the LHC7+LHC8 
data if only the statistical error is considered. The dashed red 
line in Fig.~\ref{fig:LHC8_exclusion_plot} shows the $5\sigma$ 
discovery potential for axigluons of this measurement at the LHC8 
and ${\cal L} = 25$~fb$^{-1}$.

\section{Flipped Scenario}\label{Sec:FlippedScenario}

The forward-backward asymmetry measured at the Tevatron seems to 
exceed the SM prediction while the measured charge asymmetry at 
the LHC is consistent with it. New physics scenarios like models 
including axigluons can explain the Tevatron forward-backward 
asymmetry, however, in general they also lead to a substantial 
deviation of the SM prediction for the charge asymmetry at the LHC. 

It has been suggested~\cite{Drobnak:2012cz} that this conflict 
can be solved in a ''flipped`` axigluon model where the couplings 
to up-type and down-type SM quarks is chosen differently from 
each other. At the Tevatron, the partonic $u\bar{u}$-channel 
largely dominates the cross section since both PDFs, 
$f_{q/p}(x_{1})$ and $f_{\bar q/\bar p}(x_{2})$ are of valence 
type. In contrast, at the LHC, only one of the two PDFs is a valence 
distribution. Therefore, also the ratio of contributions from the 
$u\bar{u}$-channel to the $d\bar{d}$-channel is much smaller at 
the LHC than at the Tevatron. The average value of the partonic 
momentum fractions $\tau=x_{1}x_{2}$ is smaller at the LHC. This 
would reduce the enhancement by the valence component of the PDFs, 
but there is still enough room to fit the $g_{V,A}^{u}$ and 
$g_{V,A}^{d}$ couplings in such a way that both, the Tevatron and 
LHC measurements, can be made compatible~\cite{Drobnak:2012cz}. 

\begin{figure}[t]
\includegraphics[width=7.3cm]{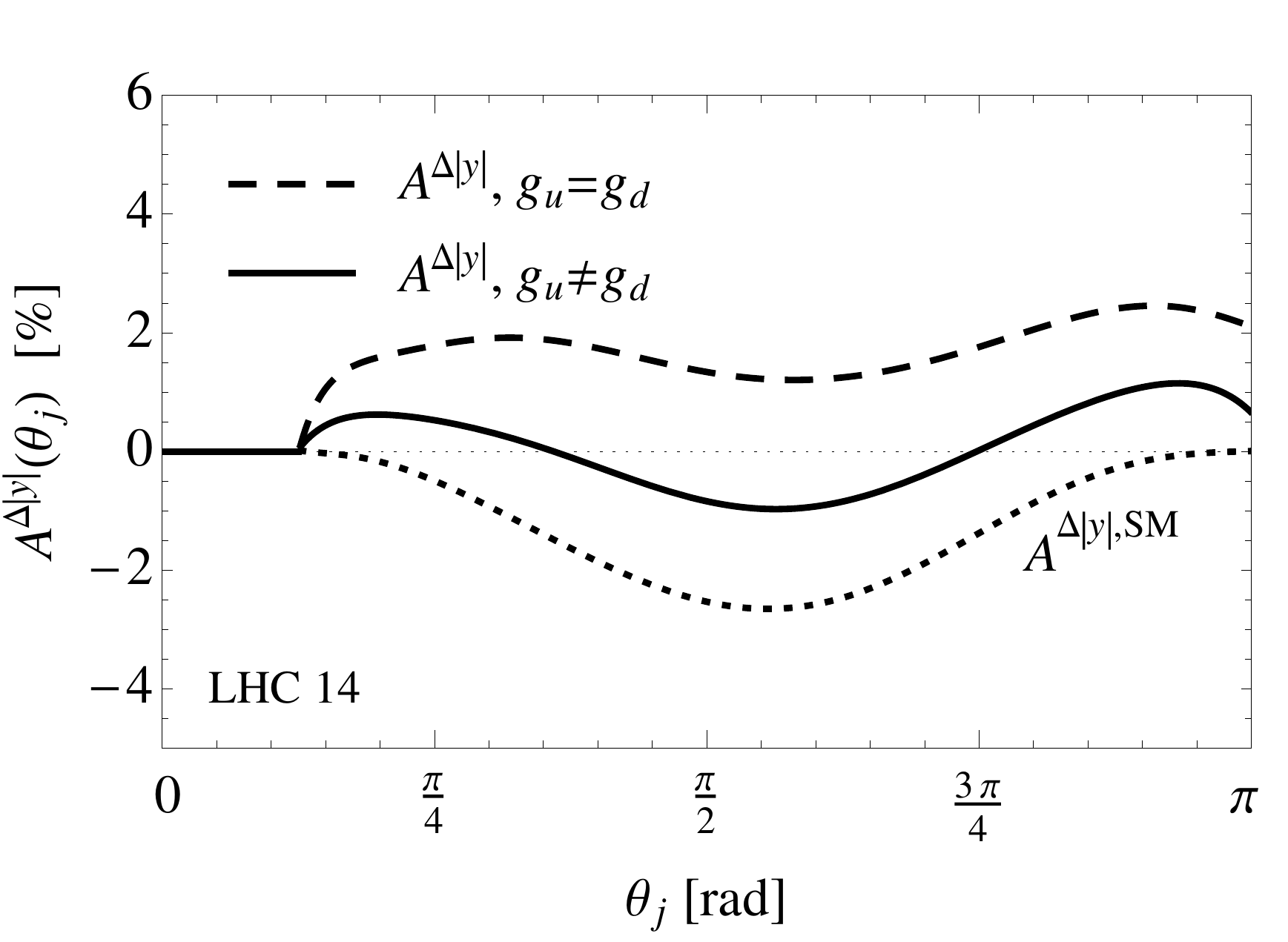}\hspace{0.5cm}
\includegraphics[width=7.3cm]{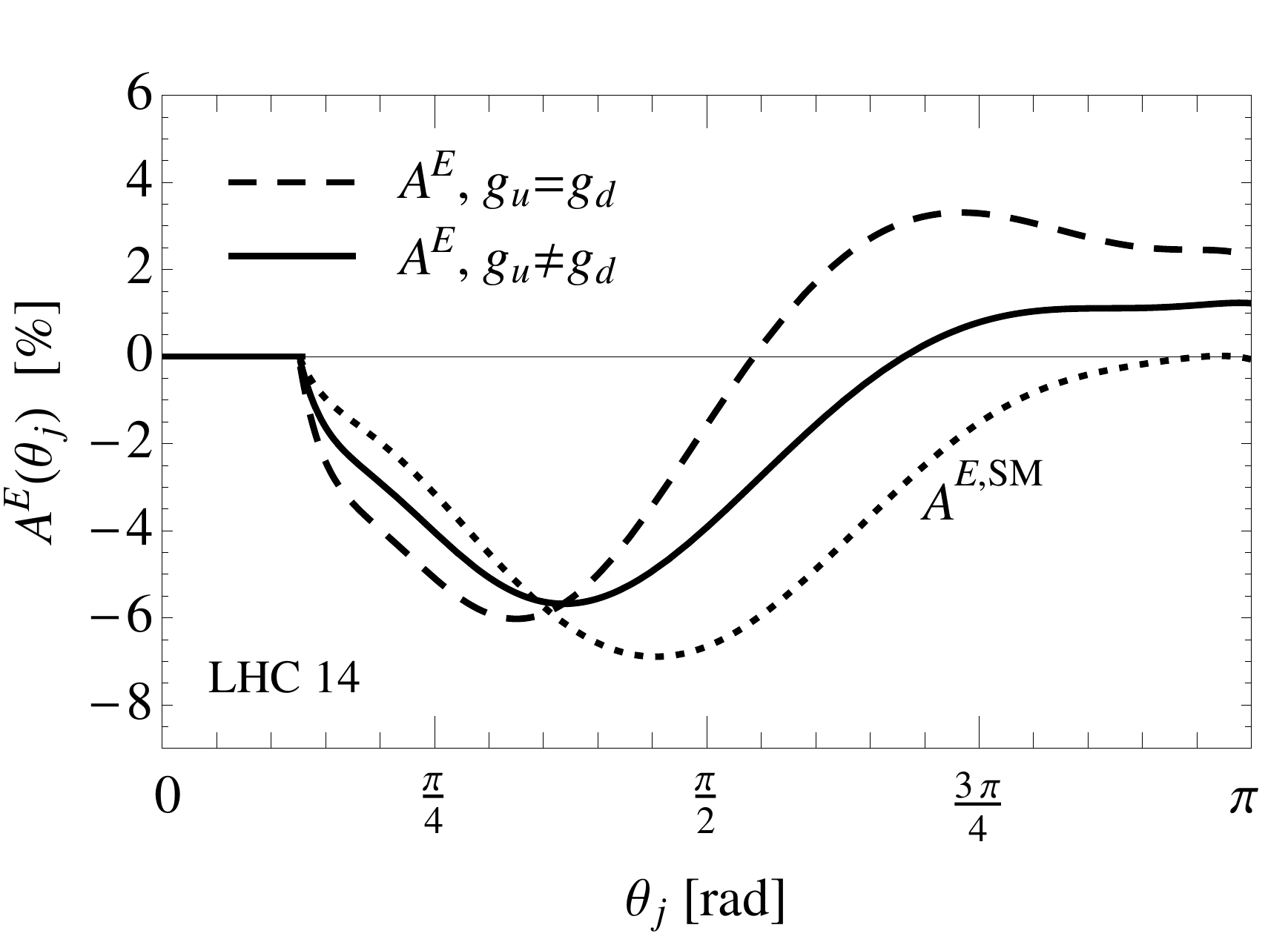}
\caption{
Rapidity and energy asymmetries in a flipped axigluon scenario at 
the LHC with $\sqrt{S} = 14$~TeV, $p_{Tj} \ge 25$~GeV and $|y_{j}| 
\le 2.5$. Dotted lines denote the SM prediction, dashed lines the 
axigluon scenario with $\tilde{g}_{U} = \tilde{g}_{D}$ and solid 
lines the flipped scenario with $\tilde{g}_{U} \ne \tilde{g}_{D}$. 
Left panel: dependence of $A^{\Delta|y|}(\theta_{j})$ on the jet 
scattering angle with $|y_{t\bar{t}j}| \ge 1$ and $|\Delta|y|| \ge 1$; 
right panel: $A^{\Delta E}(\theta_{j})$ with $|y_{t\bar{t}j}| \ge 1$ and 
$|\Delta E| \ge100$~GeV. 
\label{fig:LHC_flipped_scenario}
}
\end{figure}

We have investigated the influence of axigluons in such a model on 
the $t\bar{t}+jet$ charge asymmetries and base our numerical results 
on the fitted values of Ref.~\cite{Drobnak:2012cz}, $\tilde{g}_{Q} 
= 0.5$, $\tilde{g}_{U} = 0.32$ and $\tilde{g}_{D} = -1.2$ for $m_{A} 
= 350$~GeV and $\Gamma_{A} = 0.2 \cdot m_{A}$. Our results are shown 
in Fig.~\ref{fig:LHC_flipped_scenario}. In the left panel we show 
the rapidity asymmetry as a function of the jet scattering angle 
$\theta_{j}$ and in the right panel the energy asymmetry. 
All results are for the LHC at $\sqrt{S} = 14$~TeV, with ``detector 
cuts'' $p_{Tj} \ge 25$~GeV and $|y_{j}| \le 2.5$. The dotted 
lines correspond to the SM prediction, the dashed lines include 
axigluons with $\tilde{g}_{U} = \tilde{g}_{D} = 0.32$, while the 
solid lines show the results for the flipped axigluon scenario with 
$\tilde{g}_{U} = 0.32$ and $\tilde{g}_{D} = -1.2$. For the rapidity 
asymmetry we have applied the additional cuts $|y_{t\bar{t}j}| \ge 
1$ and $|\Delta|y|| \ge 1$ in order to enhance the observable 
asymmetry, as explained in the previous sections. The curve for 
$\tilde{g}_{U} = \tilde{g}_{D} = 0.32$ is shifted to positive values 
leading to an asymmetry difference with respect to the SM result of 
about $4\,\%$ for $\theta_{j} \simeq \pi/2$. The $\theta_{j}$ 
dependence in the flipped scenario is slightly distorted. This is 
a consequence of the fact that the hadron level result is 
a superposition of the $q\bar{q}$ and $qg$ partonic channels 
which are affected by the presence of axigluons in quite different 
ways. We note that after integration over the full $\theta_{j}$ 
range, the asymmetry is very small, $A^{\Delta|y|}_{\text{int}} 
= 0.32\,\%$. Therefore, a measurement of the dependence on the jet 
scattering angle will be very important to distinguish the flipped 
scenario from the SM or from other model predictions. 

Similarly, the energy asymmetry shown in the right panel of 
Fig.~\ref{fig:LHC_flipped_scenario} can be helpful to extract 
information about axigluon model parameters. While again the 
contribution from axigluons in a flipped scenario to the integrated 
rapidity asymmetry is almost zero, one still can observe substantial 
deviations in the $\theta_{j}$ dependence at the level of 
2 to 3\,\%.

\section{Concluding Remarks}\label{Sec:Conclusion}

In the present paper we have studied the impact of massive 
color-octet states on the charge asymmetry in $t\bar{t}+jet$ production at
the LHC. We investigated the incline and energy 
asymmetry~\cite{Berge:2013xsa} as well as the conventional rapidity 
asymmetry. The masses of the color octet states have been chosen 
in the range of relatively small values between $100$~GeV and $400$~GeV
motivated by~\cite{Gross:2012bz}. 

In a scenario with purely axial-vector couplings we found that 
large differences with respect to the SM prediction for all three 
asymmetries could be observed at the LHC with a center of mass 
energy of $14$~TeV. For each asymmetry the difference to the SM 
prediction can be enhanced by appropriate kinematical cuts 
leading to absolute differences of $10 - 20\%$. In particular we 
have shown that there are striking differences with respect to the 
SM predictions in the $\theta_j$-differential distributions. The 
SM asymmetries always tend to zero for $\theta_j \to 0, \pi$,
whereas the energy and rapidity asymmetry turn out to be finite and 
large in the SM extension with axigluons. We have shown that the 
whole parameter space of the axigluon scenario suggested in 
Ref.~\cite{Gross:2012bz} can be tested with a luminosity of about 
$65\,$fb$^{-1}$.

Also with data from LHC runs at the lower CM energy of $8$~TeV 
one should be able to derive meaningful exclusion limits on the 
axigluon coupling $\alpha_A$ and mass $m_A$. Large parts of the 
parameter space of the model suggested in Ref.~\cite{Gross:2012bz} 
could be excluded with already existing data. One should mention 
though, that our investigations have been made at the top-quark 
production level at leading order QCD. A full simulation of the 
event characteristics including the top-quark decay and next-to-leading 
order corrections requires further investigations. 

We have also considered the so-called flipped scenario suggested
in Ref.~\cite{Drobnak:2012cz} where the massive color-octet states 
have both vector and axial-vector couplings to SM quarks. The flipped 
scenario was suggested as an explanation for the large deviation of 
the $t\bar{t}$ forward-backward asymmetry measured at the Tevatron 
while keeping the deviation of observables which have already been 
measured at the LHC small. We have shown that such a scenario would 
indeed lead to only small integrated charge asymmetries in 
$t\bar{t}+jet$ final state. However, the differential asymmetries 
would show substantial deviations with respect to the SM prediction. 
Therefore, measuring the $\theta_j$-dependent asymmetries in 
$t\bar{t}+jet$ production at the LHC should be able to confirm or 
exclude such scenarios.

\section*{Acknowledgments}

We thank W.\ Bernreuther for helpful discussions and S.\ Westhoff 
for a careful reading of the manuscript.
This work was supported by the 27 Initiative and Networking Fund of
the Helmholtz Association, contract HA-101 (\textquoteleft{}Physics
at the Terascale\textquoteright{}) and by the Research Center 
\textquoteleft{}Elementary Forces and Mathematical 
Foundations\textquoteright{} of the Johannes Gutenberg-Universit\"at 
Mainz.


\bibliographystyle{apsrev}
\bibliography{bibppttj}

\begin{thebibliography}{67}
\expandafter\ifx\csname natexlab\endcsname\relax\def\natexlab#1{#1}\fi
\expandafter\ifx\csname bibnamefont\endcsname\relax
  \def\bibnamefont#1{#1}\fi
\expandafter\ifx\csname bibfnamefont\endcsname\relax
  \def\bibfnamefont#1{#1}\fi
\expandafter\ifx\csname citenamefont\endcsname\relax
  \def\citenamefont#1{#1}\fi
\expandafter\ifx\csname url\endcsname\relax
  \def\url#1{\texttt{#1}}\fi
\expandafter\ifx\csname urlprefix\endcsname\relax\def\urlprefix{URL }\fi
\providecommand{\bibinfo}[2]{#2}
\providecommand{\eprint}[2][]{\url{#2}}

\bibitem[{\citenamefont{Brown et~al.}(1979)\citenamefont{Brown, Sahdev, and
  Mikaelian}}]{Brown:1979dd}
\bibinfo{author}{\bibfnamefont{R.}~\bibnamefont{Brown}},
  \bibinfo{author}{\bibfnamefont{D.}~\bibnamefont{Sahdev}}, \bibnamefont{and}
  \bibinfo{author}{\bibfnamefont{K.}~\bibnamefont{Mikaelian}},
  \bibinfo{journal}{Phys.Rev.Lett.} \textbf{\bibinfo{volume}{43}},
  \bibinfo{pages}{1069} (\bibinfo{year}{1979}).

\bibitem[{\citenamefont{Kuhn and Rodrigo}(1998)}]{Kuhn:1998jr}
\bibinfo{author}{\bibfnamefont{J.~H.} \bibnamefont{Kuhn}} \bibnamefont{and}
  \bibinfo{author}{\bibfnamefont{G.}~\bibnamefont{Rodrigo}},
  \bibinfo{journal}{Phys.Rev.Lett.} \textbf{\bibinfo{volume}{81}},
  \bibinfo{pages}{49} (\bibinfo{year}{1998}), \eprint{arXiv:hep-ph/9802268}.

\bibitem[{\citenamefont{Kuhn and Rodrigo}(1999)}]{Kuhn:1998kw}
\bibinfo{author}{\bibfnamefont{J.~H.} \bibnamefont{Kuhn}} \bibnamefont{and}
  \bibinfo{author}{\bibfnamefont{G.}~\bibnamefont{Rodrigo}},
  \bibinfo{journal}{Phys.Rev.} \textbf{\bibinfo{volume}{D59}},
  \bibinfo{pages}{054017} (\bibinfo{year}{1999}),
  \eprint{arXiv:hep-ph/9807420}.

\bibitem[{\citenamefont{Aaltonen et~al.}(2013{\natexlab{a}})}]{Aaltonen:2012it}
\bibinfo{author}{\bibfnamefont{T.}~\bibnamefont{Aaltonen}} \bibnamefont{et~al.}
  (\bibinfo{collaboration}{CDF Collaboration}), \bibinfo{journal}{Phys.Rev.}
  \textbf{\bibinfo{volume}{D87}}, \bibinfo{pages}{092002}
  (\bibinfo{year}{2013}{\natexlab{a}}), \eprint{arXiv:1211.1003}.

\bibitem[{\citenamefont{Abazov et~al.}(2014{\natexlab{a}})}]{Abazov:2014cca}
\bibinfo{author}{\bibfnamefont{V.~M.} \bibnamefont{Abazov}}
  \bibnamefont{et~al.} (\bibinfo{collaboration}{D0 Collaboration})
  (\bibinfo{year}{2014}{\natexlab{a}}), \eprint{arXiv:1405.0421}.

\bibitem[{\citenamefont{Aaltonen
  et~al.}(2013{\natexlab{b}})}]{Aaltonen:2013vaf}
\bibinfo{author}{\bibfnamefont{T.~A.} \bibnamefont{Aaltonen}}
  \bibnamefont{et~al.} (\bibinfo{collaboration}{CDF Collaboration}),
  \bibinfo{journal}{Phys.Rev.} \textbf{\bibinfo{volume}{D88}},
  \bibinfo{pages}{072003} (\bibinfo{year}{2013}{\natexlab{b}}),
  \eprint{arXiv:1308.1120}.

\bibitem[{\citenamefont{Abazov et~al.}(2014{\natexlab{b}})}]{Abazov:2014oea}
\bibinfo{author}{\bibfnamefont{V.~M.} \bibnamefont{Abazov}}
  \bibnamefont{et~al.} (\bibinfo{collaboration}{D0 Collaboration})
  (\bibinfo{year}{2014}{\natexlab{b}}), \eprint{arXiv:1403.1294}.

\bibitem[{\citenamefont{Aaltonen et~al.}(2014)}]{Aaltonen:2014eva}
\bibinfo{author}{\bibfnamefont{T.~A.} \bibnamefont{Aaltonen}}
  \bibnamefont{et~al.} (\bibinfo{collaboration}{CDF Collaboration})
  (\bibinfo{year}{2014}), \eprint{arXiv:1404.3698}.

\bibitem[{\citenamefont{Abazov et~al.}(2013)}]{Abazov:2013wxa}
\bibinfo{author}{\bibfnamefont{V.~M.} \bibnamefont{Abazov}}
  \bibnamefont{et~al.} (\bibinfo{collaboration}{D0 Collaboration}),
  \bibinfo{journal}{Phys.Rev.} \textbf{\bibinfo{volume}{D88}},
  \bibinfo{pages}{112002} (\bibinfo{year}{2013}), \eprint{arXiv:1308.6690}.

\bibitem[{\citenamefont{Kuhn and Rodrigo}(2012)}]{Kuhn:2011ri}
\bibinfo{author}{\bibfnamefont{J.~H.} \bibnamefont{Kuhn}} \bibnamefont{and}
  \bibinfo{author}{\bibfnamefont{G.}~\bibnamefont{Rodrigo}},
  \bibinfo{journal}{JHEP} \textbf{\bibinfo{volume}{1201}}, \bibinfo{pages}{063}
  (\bibinfo{year}{2012}), \eprint{arXiv:1109.6830}.

\bibitem[{\citenamefont{Bernreuther and Si}(2012)}]{Bernreuther:2012sx}
\bibinfo{author}{\bibfnamefont{W.}~\bibnamefont{Bernreuther}} \bibnamefont{and}
  \bibinfo{author}{\bibfnamefont{Z.-G.} \bibnamefont{Si}},
  \bibinfo{journal}{Phys.Rev.} \textbf{\bibinfo{volume}{D86}},
  \bibinfo{pages}{034026} (\bibinfo{year}{2012}), \eprint{arXiv:1205.6580}.

\bibitem[{\citenamefont{Hollik and Pagani}(2011)}]{Hollik:2011ps}
\bibinfo{author}{\bibfnamefont{W.}~\bibnamefont{Hollik}} \bibnamefont{and}
  \bibinfo{author}{\bibfnamefont{D.}~\bibnamefont{Pagani}},
  \bibinfo{journal}{Phys.Rev.} \textbf{\bibinfo{volume}{D84}},
  \bibinfo{pages}{093003} (\bibinfo{year}{2011}), \eprint{1107.2606}.

\bibitem[{\citenamefont{Aad et~al.}(2014)}]{Aad:2013cea}
\bibinfo{author}{\bibfnamefont{G.}~\bibnamefont{Aad}} \bibnamefont{et~al.}
  (\bibinfo{collaboration}{ATLAS Collaboration}), \bibinfo{journal}{JHEP}
  \textbf{\bibinfo{volume}{1402}}, \bibinfo{pages}{107} (\bibinfo{year}{2014}),
  \eprint{1311.6724}.

\bibitem[{\citenamefont{Chatrchyan et~al.}(2012)}]{Chatrchyan:2012cxa}
\bibinfo{author}{\bibfnamefont{S.}~\bibnamefont{Chatrchyan}}
  \bibnamefont{et~al.} (\bibinfo{collaboration}{CMS Collaboration}),
  \bibinfo{journal}{Phys.Lett.} \textbf{\bibinfo{volume}{B717}},
  \bibinfo{pages}{129} (\bibinfo{year}{2012}), \eprint{arXiv:1207.0065}.

\bibitem[{\citenamefont{{CMS Collaboration}}(2013)}]{CMS:2013nfa}
\bibinfo{author}{\bibnamefont{{CMS Collaboration}}},
  \bibinfo{journal}{CMS-PAS-TOP-12-033}  (\bibinfo{year}{2013}).

\bibitem[{\citenamefont{{ATLAS
  Collaboration}}(2012{\natexlab{a}})}]{ATLAS:2012sla}
\bibinfo{author}{\bibnamefont{{ATLAS Collaboration}}},
  \bibinfo{journal}{ATLAS-CONF-2012-057, ATLAS-COM-CONF-2012-060}
  (\bibinfo{year}{2012}{\natexlab{a}}).

\bibitem[{\citenamefont{Chatrchyan et~al.}(2014)}]{Chatrchyan:2014yta}
\bibinfo{author}{\bibfnamefont{S.}~\bibnamefont{Chatrchyan}}
  \bibnamefont{et~al.} (\bibinfo{collaboration}{CMS Collaboration}),
  \bibinfo{journal}{JHEP} \textbf{\bibinfo{volume}{1404}}, \bibinfo{pages}{191}
  (\bibinfo{year}{2014}), \eprint{1402.3803}.

\bibitem[{\citenamefont{{ATLAS Collaboration, CMS
  Collaboration}}(2014)}]{CMS-PAS-TOP-14-006}
\bibinfo{author}{\bibnamefont{{ATLAS Collaboration, CMS Collaboration}}},
  \bibinfo{type}{Tech. Rep.} \bibinfo{number}{ATLAS-CONF-2014-012,
  ATLAS-COM-CONF-2014-014, CMS-PAS-TOP-14-006}, \bibinfo{institution}{CERN},
  \bibinfo{address}{Geneva} (\bibinfo{year}{2014}).

\bibitem[{\citenamefont{Halzen et~al.}(1987)\citenamefont{Halzen, Hoyer, and
  Kim}}]{Halzen:1987xd}
\bibinfo{author}{\bibfnamefont{F.}~\bibnamefont{Halzen}},
  \bibinfo{author}{\bibfnamefont{P.}~\bibnamefont{Hoyer}}, \bibnamefont{and}
  \bibinfo{author}{\bibfnamefont{C.}~\bibnamefont{Kim}},
  \bibinfo{journal}{Phys.Lett.} \textbf{\bibinfo{volume}{B195}},
  \bibinfo{pages}{74} (\bibinfo{year}{1987}).

\bibitem[{\citenamefont{Dittmaier et~al.}(2007)\citenamefont{Dittmaier, Uwer,
  and Weinzierl}}]{Dittmaier:2007wz}
\bibinfo{author}{\bibfnamefont{S.}~\bibnamefont{Dittmaier}},
  \bibinfo{author}{\bibfnamefont{P.}~\bibnamefont{Uwer}}, \bibnamefont{and}
  \bibinfo{author}{\bibfnamefont{S.}~\bibnamefont{Weinzierl}},
  \bibinfo{journal}{Phys.Rev.Lett.} \textbf{\bibinfo{volume}{98}},
  \bibinfo{pages}{262002} (\bibinfo{year}{2007}),
  \eprint{arXiv:hep-ph/0703120}.

\bibitem[{\citenamefont{Dittmaier et~al.}(2009)\citenamefont{Dittmaier, Uwer,
  and Weinzierl}}]{Dittmaier:2008uj}
\bibinfo{author}{\bibfnamefont{S.}~\bibnamefont{Dittmaier}},
  \bibinfo{author}{\bibfnamefont{P.}~\bibnamefont{Uwer}}, \bibnamefont{and}
  \bibinfo{author}{\bibfnamefont{S.}~\bibnamefont{Weinzierl}},
  \bibinfo{journal}{Eur.Phys.J.} \textbf{\bibinfo{volume}{C59}},
  \bibinfo{pages}{625} (\bibinfo{year}{2009}), \eprint{arXiv:0810.0452}.

\bibitem[{\citenamefont{Melnikov and Schulze}(2010)}]{Melnikov:2010iu}
\bibinfo{author}{\bibfnamefont{K.}~\bibnamefont{Melnikov}} \bibnamefont{and}
  \bibinfo{author}{\bibfnamefont{M.}~\bibnamefont{Schulze}},
  \bibinfo{journal}{Nucl.Phys.} \textbf{\bibinfo{volume}{B840}},
  \bibinfo{pages}{129} (\bibinfo{year}{2010}), \eprint{arXiv:1004.3284}.

\bibitem[{\citenamefont{Ahrens et~al.}(2011)\citenamefont{Ahrens, Ferroglia,
  Neubert, Pecjak, and Yang}}]{Ahrens:2011uf}
\bibinfo{author}{\bibfnamefont{V.}~\bibnamefont{Ahrens}},
  \bibinfo{author}{\bibfnamefont{A.}~\bibnamefont{Ferroglia}},
  \bibinfo{author}{\bibfnamefont{M.}~\bibnamefont{Neubert}},
  \bibinfo{author}{\bibfnamefont{B.~D.} \bibnamefont{Pecjak}},
  \bibnamefont{and} \bibinfo{author}{\bibfnamefont{L.~L.} \bibnamefont{Yang}},
  \bibinfo{journal}{Phys.Rev.} \textbf{\bibinfo{volume}{D84}},
  \bibinfo{pages}{074004} (\bibinfo{year}{2011}), \eprint{arXiv:1106.6051}.

\bibitem[{\citenamefont{Alioli et~al.}(2012)\citenamefont{Alioli, Moch, and
  Uwer}}]{Alioli:2011as}
\bibinfo{author}{\bibfnamefont{S.}~\bibnamefont{Alioli}},
  \bibinfo{author}{\bibfnamefont{S.-O.} \bibnamefont{Moch}}, \bibnamefont{and}
  \bibinfo{author}{\bibfnamefont{P.}~\bibnamefont{Uwer}},
  \bibinfo{journal}{JHEP} \textbf{\bibinfo{volume}{1201}}, \bibinfo{pages}{137}
  (\bibinfo{year}{2012}), \eprint{arXiv:1110.5251}.

\bibitem[{\citenamefont{Melnikov et~al.}(2012)\citenamefont{Melnikov, Scharf,
  and Schulze}}]{Melnikov:2011qx}
\bibinfo{author}{\bibfnamefont{K.}~\bibnamefont{Melnikov}},
  \bibinfo{author}{\bibfnamefont{A.}~\bibnamefont{Scharf}}, \bibnamefont{and}
  \bibinfo{author}{\bibfnamefont{M.}~\bibnamefont{Schulze}},
  \bibinfo{journal}{Phys.Rev.} \textbf{\bibinfo{volume}{D85}},
  \bibinfo{pages}{054002} (\bibinfo{year}{2012}), \eprint{arXiv:1111.4991}.

\bibitem[{\citenamefont{Kardos et~al.}(2011)\citenamefont{Kardos, Papadopoulos,
  and Trocsanyi}}]{Kardos:2011qa}
\bibinfo{author}{\bibfnamefont{A.}~\bibnamefont{Kardos}},
  \bibinfo{author}{\bibfnamefont{C.}~\bibnamefont{Papadopoulos}},
  \bibnamefont{and}
  \bibinfo{author}{\bibfnamefont{Z.}~\bibnamefont{Trocsanyi}},
  \bibinfo{journal}{Phys.Lett.} \textbf{\bibinfo{volume}{B705}},
  \bibinfo{pages}{76} (\bibinfo{year}{2011}), \eprint{arXiv:1101.2672}.

\bibitem[{\citenamefont{Hoeche et~al.}(2013)\citenamefont{Hoeche, Huang,
  Luisoni, Schoenherr, and Winter}}]{Hoeche:2013mua}
\bibinfo{author}{\bibfnamefont{S.}~\bibnamefont{Hoeche}},
  \bibinfo{author}{\bibfnamefont{J.}~\bibnamefont{Huang}},
  \bibinfo{author}{\bibfnamefont{G.}~\bibnamefont{Luisoni}},
  \bibinfo{author}{\bibfnamefont{M.}~\bibnamefont{Schoenherr}},
  \bibnamefont{and} \bibinfo{author}{\bibfnamefont{J.}~\bibnamefont{Winter}},
  \bibinfo{journal}{Phys.Rev.} \textbf{\bibinfo{volume}{D88}},
  \bibinfo{pages}{014040} (\bibinfo{year}{2013}), \eprint{arXiv:1306.2703}.

\bibitem[{\citenamefont{Westhoff}(2013)}]{Westhoff:2013ixa}
\bibinfo{author}{\bibfnamefont{S.}~\bibnamefont{Westhoff}}
  (\bibinfo{year}{2013}), \eprint{arXiv:1311.1127}.

\bibitem[{\citenamefont{Aguilar-Saavedra and
  Perez-Victoria}(2013)}]{Aguilar-Saavedra:2013rza}
\bibinfo{author}{\bibfnamefont{J.}~\bibnamefont{Aguilar-Saavedra}}
  \bibnamefont{and}
  \bibinfo{author}{\bibfnamefont{M.}~\bibnamefont{Perez-Victoria}},
  \bibinfo{journal}{J.Phys.Conf.Ser.} \textbf{\bibinfo{volume}{447}},
  \bibinfo{pages}{012015} (\bibinfo{year}{2013}), \eprint{1302.6618}.

\bibitem[{\citenamefont{Berger}(2013)}]{Berger:2013et}
\bibinfo{author}{\bibfnamefont{E.~L.} \bibnamefont{Berger}}
  (\bibinfo{year}{2013}), \eprint{1301.5053}.

\bibitem[{\citenamefont{Aguilar-Saavedra and
  Perez-Victoria}(2011{\natexlab{a}})}]{AguilarSaavedra:2011vw}
\bibinfo{author}{\bibfnamefont{J.}~\bibnamefont{Aguilar-Saavedra}}
  \bibnamefont{and}
  \bibinfo{author}{\bibfnamefont{M.}~\bibnamefont{Perez-Victoria}},
  \bibinfo{journal}{JHEP} \textbf{\bibinfo{volume}{1105}}, \bibinfo{pages}{034}
  (\bibinfo{year}{2011}{\natexlab{a}}), \eprint{1103.2765}.

\bibitem[{\citenamefont{Kamenik et~al.}(2012)\citenamefont{Kamenik, Shu, and
  Zupan}}]{Kamenik:2011wt}
\bibinfo{author}{\bibfnamefont{J.~F.} \bibnamefont{Kamenik}},
  \bibinfo{author}{\bibfnamefont{J.}~\bibnamefont{Shu}}, \bibnamefont{and}
  \bibinfo{author}{\bibfnamefont{J.}~\bibnamefont{Zupan}},
  \bibinfo{journal}{Eur.Phys.J.} \textbf{\bibinfo{volume}{C72}},
  \bibinfo{pages}{2102} (\bibinfo{year}{2012}), \eprint{arXiv:1107.5257}.

\bibitem[{\citenamefont{Frampton and Glashow}(1987)}]{Frampton:1987dn}
\bibinfo{author}{\bibfnamefont{P.~H.} \bibnamefont{Frampton}} \bibnamefont{and}
  \bibinfo{author}{\bibfnamefont{S.~L.} \bibnamefont{Glashow}},
  \bibinfo{journal}{Phys.Lett.} \textbf{\bibinfo{volume}{B190}},
  \bibinfo{pages}{157} (\bibinfo{year}{1987}).

\bibitem[{\citenamefont{Bagger et~al.}(1988)\citenamefont{Bagger, Schmidt, and
  King}}]{Bagger:1987fz}
\bibinfo{author}{\bibfnamefont{J.}~\bibnamefont{Bagger}},
  \bibinfo{author}{\bibfnamefont{C.}~\bibnamefont{Schmidt}}, \bibnamefont{and}
  \bibinfo{author}{\bibfnamefont{S.}~\bibnamefont{King}},
  \bibinfo{journal}{Phys.Rev.} \textbf{\bibinfo{volume}{D37}},
  \bibinfo{pages}{1188} (\bibinfo{year}{1988}).

\bibitem[{\citenamefont{Chivukula et~al.}(2010)\citenamefont{Chivukula,
  Simmons, and Yuan}}]{Chivukula:2010fk}
\bibinfo{author}{\bibfnamefont{R.~S.} \bibnamefont{Chivukula}},
  \bibinfo{author}{\bibfnamefont{E.~H.} \bibnamefont{Simmons}},
  \bibnamefont{and} \bibinfo{author}{\bibfnamefont{C.-P.} \bibnamefont{Yuan}},
  \bibinfo{journal}{Phys.Rev.} \textbf{\bibinfo{volume}{D82}},
  \bibinfo{pages}{094009} (\bibinfo{year}{2010}), \eprint{arXiv:1007.0260}.

\bibitem[{\citenamefont{Díaz and Zerwekh}(2013)}]{Diaz:2013tfa}
\bibinfo{author}{\bibfnamefont{B.}~\bibnamefont{Díaz}} \bibnamefont{and}
  \bibinfo{author}{\bibfnamefont{A.~R.} \bibnamefont{Zerwekh}},
  \bibinfo{journal}{Int.J.Mod.Phys.} \textbf{\bibinfo{volume}{A28}},
  \bibinfo{pages}{1350133} (\bibinfo{year}{2013}), \eprint{arXiv:1308.0166}.

\bibitem[{\citenamefont{Haisch and Westhoff}(2011)}]{Haisch:2011up}
\bibinfo{author}{\bibfnamefont{U.}~\bibnamefont{Haisch}} \bibnamefont{and}
  \bibinfo{author}{\bibfnamefont{S.}~\bibnamefont{Westhoff}},
  \bibinfo{journal}{JHEP} \textbf{\bibinfo{volume}{1108}}, \bibinfo{pages}{088}
  (\bibinfo{year}{2011}), \eprint{1106.0529}.

\bibitem[{\citenamefont{Xiao et~al.}(2010)\citenamefont{Xiao, Wang, and
  Zhu}}]{Xiao:2010ph}
\bibinfo{author}{\bibfnamefont{B.}~\bibnamefont{Xiao}},
  \bibinfo{author}{\bibfnamefont{Y.-k.} \bibnamefont{Wang}}, \bibnamefont{and}
  \bibinfo{author}{\bibfnamefont{S.-h.} \bibnamefont{Zhu}}
  (\bibinfo{year}{2010}), \eprint{1011.0152}.

\bibitem[{\citenamefont{Marques~Tavares and Schmaltz}(2011)}]{Tavares:2011zg}
\bibinfo{author}{\bibfnamefont{G.}~\bibnamefont{Marques~Tavares}}
  \bibnamefont{and} \bibinfo{author}{\bibfnamefont{M.}~\bibnamefont{Schmaltz}},
  \bibinfo{journal}{Phys.Rev.} \textbf{\bibinfo{volume}{D84}},
  \bibinfo{pages}{054008} (\bibinfo{year}{2011}), \eprint{arXiv:1107.0978}.

\bibitem[{\citenamefont{Barcelo et~al.}(2012)\citenamefont{Barcelo, Carmona,
  Masip, and Santiago}}]{Barcelo:2011vk}
\bibinfo{author}{\bibfnamefont{R.}~\bibnamefont{Barcelo}},
  \bibinfo{author}{\bibfnamefont{A.}~\bibnamefont{Carmona}},
  \bibinfo{author}{\bibfnamefont{M.}~\bibnamefont{Masip}}, \bibnamefont{and}
  \bibinfo{author}{\bibfnamefont{J.}~\bibnamefont{Santiago}},
  \bibinfo{journal}{Phys.Lett.} \textbf{\bibinfo{volume}{B707}},
  \bibinfo{pages}{88} (\bibinfo{year}{2012}), \eprint{1106.4054}.

\bibitem[{\citenamefont{Alvarez et~al.}(2011)\citenamefont{Alvarez, Da~Rold,
  Vietto, and Szynkman}}]{Alvarez:2011hi}
\bibinfo{author}{\bibfnamefont{E.}~\bibnamefont{Alvarez}},
  \bibinfo{author}{\bibfnamefont{L.}~\bibnamefont{Da~Rold}},
  \bibinfo{author}{\bibfnamefont{J.~I.~S.} \bibnamefont{Vietto}},
  \bibnamefont{and} \bibinfo{author}{\bibfnamefont{A.}~\bibnamefont{Szynkman}},
  \bibinfo{journal}{JHEP} \textbf{\bibinfo{volume}{1109}}, \bibinfo{pages}{007}
  (\bibinfo{year}{2011}), \eprint{1107.1473}.

\bibitem[{\citenamefont{Aguilar-Saavedra and
  Perez-Victoria}(2011{\natexlab{b}})}]{AguilarSaavedra:2011ci}
\bibinfo{author}{\bibfnamefont{J.}~\bibnamefont{Aguilar-Saavedra}}
  \bibnamefont{and}
  \bibinfo{author}{\bibfnamefont{M.}~\bibnamefont{Perez-Victoria}},
  \bibinfo{journal}{Phys.Lett.} \textbf{\bibinfo{volume}{B705}},
  \bibinfo{pages}{228} (\bibinfo{year}{2011}{\natexlab{b}}),
  \eprint{1107.2120}.

\bibitem[{\citenamefont{Krnjaic}(2012)}]{Krnjaic:2011ub}
\bibinfo{author}{\bibfnamefont{G.~Z.} \bibnamefont{Krnjaic}},
  \bibinfo{journal}{Phys.Rev.} \textbf{\bibinfo{volume}{D85}},
  \bibinfo{pages}{014030} (\bibinfo{year}{2012}), \eprint{arXiv:1109.0648}.

\bibitem[{\citenamefont{Aguilar-Saavedra
  et~al.}(2014)\citenamefont{Aguilar-Saavedra, Álvarez, Juste, and
  Rubbo}}]{Aguilar-Saavedra:2014vta}
\bibinfo{author}{\bibfnamefont{J.}~\bibnamefont{Aguilar-Saavedra}},
  \bibinfo{author}{\bibfnamefont{E.}~\bibnamefont{Álvarez}},
  \bibinfo{author}{\bibfnamefont{A.}~\bibnamefont{Juste}}, \bibnamefont{and}
  \bibinfo{author}{\bibfnamefont{F.}~\bibnamefont{Rubbo}},
  \bibinfo{journal}{JHEP} \textbf{\bibinfo{volume}{1404}}, \bibinfo{pages}{188}
  (\bibinfo{year}{2014}), \eprint{arXiv:1402.3598}.

\bibitem[{\citenamefont{Tianjun et~al.}(2013)\citenamefont{Tianjun, Xia,
  You-kai, and Shou-hua}}]{Tianjun:2013joa}
\bibinfo{author}{\bibfnamefont{L.}~\bibnamefont{Tianjun}},
  \bibinfo{author}{\bibfnamefont{W.}~\bibnamefont{Xia}},
  \bibinfo{author}{\bibfnamefont{W.}~\bibnamefont{You-kai}}, \bibnamefont{and}
  \bibinfo{author}{\bibfnamefont{Z.}~\bibnamefont{Shou-hua}}
  (\bibinfo{year}{2013}), \eprint{1306.3586}.

\bibitem[{\citenamefont{Carmona et~al.}(2014)\citenamefont{Carmona, Chala,
  Falkowski, Khatibi, Najafabadi et~al.}}]{Carmona:2014gra}
\bibinfo{author}{\bibfnamefont{A.}~\bibnamefont{Carmona}},
  \bibinfo{author}{\bibfnamefont{M.}~\bibnamefont{Chala}},
  \bibinfo{author}{\bibfnamefont{A.}~\bibnamefont{Falkowski}},
  \bibinfo{author}{\bibfnamefont{S.}~\bibnamefont{Khatibi}},
  \bibinfo{author}{\bibfnamefont{M.~M.} \bibnamefont{Najafabadi}},
  \bibnamefont{et~al.} (\bibinfo{year}{2014}), \eprint{arXiv:1401.2443}.

\bibitem[{\citenamefont{Aguilar-Saavedra}(2014{\natexlab{a}})}]{Aguilar-Saavedra:2014yea}
\bibinfo{author}{\bibfnamefont{J.}~\bibnamefont{Aguilar-Saavedra}}
  (\bibinfo{year}{2014}{\natexlab{a}}), \eprint{1405.1412}.

\bibitem[{\citenamefont{Jung et~al.}(2014)\citenamefont{Jung, Ko, Yoon, and
  Yu}}]{Jung:2014gfa}
\bibinfo{author}{\bibfnamefont{S.}~\bibnamefont{Jung}},
  \bibinfo{author}{\bibfnamefont{P.}~\bibnamefont{Ko}},
  \bibinfo{author}{\bibfnamefont{Y.~W.} \bibnamefont{Yoon}}, \bibnamefont{and}
  \bibinfo{author}{\bibfnamefont{C.}~\bibnamefont{Yu}} (\bibinfo{year}{2014}),
  \eprint{1405.5313}.

\bibitem[{\citenamefont{Ipek}(2013)}]{Ipek:2013zi}
\bibinfo{author}{\bibfnamefont{S.}~\bibnamefont{Ipek}},
  \bibinfo{journal}{Phys.Rev.} \textbf{\bibinfo{volume}{D87}},
  \bibinfo{pages}{116010} (\bibinfo{year}{2013}), \eprint{1301.3990}.

\bibitem[{\citenamefont{Falkowski et~al.}(2013)\citenamefont{Falkowski,
  Mangano, Martin, Perez, and Winter}}]{Falkowski:2012cu}
\bibinfo{author}{\bibfnamefont{A.}~\bibnamefont{Falkowski}},
  \bibinfo{author}{\bibfnamefont{M.~L.} \bibnamefont{Mangano}},
  \bibinfo{author}{\bibfnamefont{A.}~\bibnamefont{Martin}},
  \bibinfo{author}{\bibfnamefont{G.}~\bibnamefont{Perez}}, \bibnamefont{and}
  \bibinfo{author}{\bibfnamefont{J.}~\bibnamefont{Winter}},
  \bibinfo{journal}{Phys.Rev.} \textbf{\bibinfo{volume}{D87}},
  \bibinfo{pages}{034039} (\bibinfo{year}{2013}), \eprint{1212.4003}.

\bibitem[{\citenamefont{Baumgart and
  Tweedie}(2013{\natexlab{a}})}]{Baumgart:2013yra}
\bibinfo{author}{\bibfnamefont{M.}~\bibnamefont{Baumgart}} \bibnamefont{and}
  \bibinfo{author}{\bibfnamefont{B.}~\bibnamefont{Tweedie}},
  \bibinfo{journal}{JHEP} \textbf{\bibinfo{volume}{1308}}, \bibinfo{pages}{072}
  (\bibinfo{year}{2013}{\natexlab{a}}), \eprint{1303.1200}.

\bibitem[{\citenamefont{Grinstein and Murphy}(2013)}]{Grinstein:2013mia}
\bibinfo{author}{\bibfnamefont{B.}~\bibnamefont{Grinstein}} \bibnamefont{and}
  \bibinfo{author}{\bibfnamefont{C.~W.} \bibnamefont{Murphy}},
  \bibinfo{journal}{Phys.Rev.Lett.} \textbf{\bibinfo{volume}{111}},
  \bibinfo{pages}{062003} (\bibinfo{year}{2013}), \eprint{1302.6995}.

\bibitem[{\citenamefont{Baumgart and
  Tweedie}(2013{\natexlab{b}})}]{Baumgart:2012ay}
\bibinfo{author}{\bibfnamefont{M.}~\bibnamefont{Baumgart}} \bibnamefont{and}
  \bibinfo{author}{\bibfnamefont{B.}~\bibnamefont{Tweedie}},
  \bibinfo{journal}{JHEP} \textbf{\bibinfo{volume}{1303}}, \bibinfo{pages}{117}
  (\bibinfo{year}{2013}{\natexlab{b}}), \eprint{1212.4888}.

\bibitem[{\citenamefont{Aguilar-Saavedra}(2014{\natexlab{b}})}]{Aguilar-Saavedra:2014nja}
\bibinfo{author}{\bibfnamefont{J.}~\bibnamefont{Aguilar-Saavedra}}
  (\bibinfo{year}{2014}{\natexlab{b}}), \eprint{arXiv:1405.5826}.

\bibitem[{\citenamefont{Gresham et~al.}(2013)\citenamefont{Gresham, Shelton,
  and Zurek}}]{Gresham:2012kv}
\bibinfo{author}{\bibfnamefont{M.}~\bibnamefont{Gresham}},
  \bibinfo{author}{\bibfnamefont{J.}~\bibnamefont{Shelton}}, \bibnamefont{and}
  \bibinfo{author}{\bibfnamefont{K.~M.} \bibnamefont{Zurek}},
  \bibinfo{journal}{JHEP} \textbf{\bibinfo{volume}{1303}}, \bibinfo{pages}{008}
  (\bibinfo{year}{2013}), \eprint{arXiv:1212.1718}.

\bibitem[{\citenamefont{Yue et~al.}(2014)\citenamefont{Yue, Cao, and
  Zeng}}]{Yue:2014hba}
\bibinfo{author}{\bibfnamefont{C.-X.} \bibnamefont{Yue}},
  \bibinfo{author}{\bibfnamefont{S.-Y.} \bibnamefont{Cao}}, \bibnamefont{and}
  \bibinfo{author}{\bibfnamefont{Q.-G.} \bibnamefont{Zeng}},
  \bibinfo{journal}{JHEP} \textbf{\bibinfo{volume}{1404}}, \bibinfo{pages}{170}
  (\bibinfo{year}{2014}), \eprint{arXiv:1401.5159}.

\bibitem[{\citenamefont{Aaltonen
  et~al.}(2013{\natexlab{c}})}]{Aaltonen:2013hya}
\bibinfo{author}{\bibfnamefont{T.}~\bibnamefont{Aaltonen}} \bibnamefont{et~al.}
  (\bibinfo{collaboration}{CDF}), \bibinfo{journal}{Phys.Rev.Lett.}
  \textbf{\bibinfo{volume}{111}}, \bibinfo{pages}{031802}
  (\bibinfo{year}{2013}{\natexlab{c}}), \eprint{1303.2699}.

\bibitem[{\citenamefont{Dobrescu and Yu}(2013)}]{Dobrescu:2013cmh}
\bibinfo{author}{\bibfnamefont{B.~A.} \bibnamefont{Dobrescu}} \bibnamefont{and}
  \bibinfo{author}{\bibfnamefont{F.}~\bibnamefont{Yu}},
  \bibinfo{journal}{Phys.Rev.} \textbf{\bibinfo{volume}{D88}},
  \bibinfo{pages}{035021} (\bibinfo{year}{2013}), \eprint{arXiv:1306.2629}.

\bibitem[{\citenamefont{Chen et~al.}(1991)\citenamefont{Chen, Chen, and
  Zhang}}]{Chen:1991tx}
\bibinfo{author}{\bibfnamefont{L.-S.} \bibnamefont{Chen}},
  \bibinfo{author}{\bibfnamefont{Z.-J.} \bibnamefont{Chen}}, \bibnamefont{and}
  \bibinfo{author}{\bibfnamefont{J.-Z.} \bibnamefont{Zhang}},
  \bibinfo{journal}{J.Phys.} \textbf{\bibinfo{volume}{G17}},
  \bibinfo{pages}{237} (\bibinfo{year}{1991}).

\bibitem[{\citenamefont{Gross et~al.}(2013)\citenamefont{Gross,
  Marques~Tavares, Schmaltz, and Spethmann}}]{Gross:2012bz}
\bibinfo{author}{\bibfnamefont{C.}~\bibnamefont{Gross}},
  \bibinfo{author}{\bibfnamefont{G.}~\bibnamefont{Marques~Tavares}},
  \bibinfo{author}{\bibfnamefont{M.}~\bibnamefont{Schmaltz}}, \bibnamefont{and}
  \bibinfo{author}{\bibfnamefont{C.}~\bibnamefont{Spethmann}},
  \bibinfo{journal}{Phys.Rev.} \textbf{\bibinfo{volume}{D87}},
  \bibinfo{pages}{014004} (\bibinfo{year}{2013}), \eprint{arXiv:1209.6375}.

\bibitem[{\citenamefont{Berge and Westhoff}(2013)}]{Berge:2013xsa}
\bibinfo{author}{\bibfnamefont{S.}~\bibnamefont{Berge}} \bibnamefont{and}
  \bibinfo{author}{\bibfnamefont{S.}~\bibnamefont{Westhoff}},
  \bibinfo{journal}{JHEP} \textbf{\bibinfo{volume}{1307}}, \bibinfo{pages}{179}
  (\bibinfo{year}{2013}), \eprint{arXiv:1305.3272}.

\bibitem[{\citenamefont{Drobnak et~al.}(2012)\citenamefont{Drobnak, Kamenik,
  and Zupan}}]{Drobnak:2012cz}
\bibinfo{author}{\bibfnamefont{J.}~\bibnamefont{Drobnak}},
  \bibinfo{author}{\bibfnamefont{J.~F.} \bibnamefont{Kamenik}},
  \bibnamefont{and} \bibinfo{author}{\bibfnamefont{J.}~\bibnamefont{Zupan}},
  \bibinfo{journal}{Phys.Rev.} \textbf{\bibinfo{volume}{D86}},
  \bibinfo{pages}{054022} (\bibinfo{year}{2012}), \eprint{arXiv:1205.4721}.

\bibitem[{\citenamefont{Berge and Westhoff}(2012)}]{Berge:2012rc}
\bibinfo{author}{\bibfnamefont{S.}~\bibnamefont{Berge}} \bibnamefont{and}
  \bibinfo{author}{\bibfnamefont{S.}~\bibnamefont{Westhoff}},
  \bibinfo{journal}{Phys.Rev.} \textbf{\bibinfo{volume}{D86}},
  \bibinfo{pages}{094036} (\bibinfo{year}{2012}), \eprint{arXiv:1208.4104}.

\bibitem[{\citenamefont{Bernreuther et~al.}(2004)\citenamefont{Bernreuther,
  Brandenburg, Si, and Uwer}}]{Bernreuther:2004jv}
\bibinfo{author}{\bibfnamefont{W.}~\bibnamefont{Bernreuther}},
  \bibinfo{author}{\bibfnamefont{A.}~\bibnamefont{Brandenburg}},
  \bibinfo{author}{\bibfnamefont{Z.}~\bibnamefont{Si}}, \bibnamefont{and}
  \bibinfo{author}{\bibfnamefont{P.}~\bibnamefont{Uwer}},
  \bibinfo{journal}{Nucl.Phys.} \textbf{\bibinfo{volume}{B690}},
  \bibinfo{pages}{81} (\bibinfo{year}{2004}), \eprint{arXiv:hep-ph/0403035}.

\bibitem[{\citenamefont{Beringer et~al.}(2012)}]{Beringer:1900zz}
\bibinfo{author}{\bibfnamefont{J.}~\bibnamefont{Beringer}} \bibnamefont{et~al.}
  (\bibinfo{collaboration}{Particle Data Group}), \bibinfo{journal}{Phys.Rev.}
  \textbf{\bibinfo{volume}{D86}}, \bibinfo{pages}{010001}
  (\bibinfo{year}{2012}).

\bibitem[{\citenamefont{Pumplin et~al.}(2002)\citenamefont{Pumplin, Stump,
  Huston, Lai, Nadolsky et~al.}}]{Pumplin:2002vw}
\bibinfo{author}{\bibfnamefont{J.}~\bibnamefont{Pumplin}},
  \bibinfo{author}{\bibfnamefont{D.}~\bibnamefont{Stump}},
  \bibinfo{author}{\bibfnamefont{J.}~\bibnamefont{Huston}},
  \bibinfo{author}{\bibfnamefont{H.}~\bibnamefont{Lai}},
  \bibinfo{author}{\bibfnamefont{P.~M.} \bibnamefont{Nadolsky}},
  \bibnamefont{et~al.}, \bibinfo{journal}{JHEP}
  \textbf{\bibinfo{volume}{0207}}, \bibinfo{pages}{012} (\bibinfo{year}{2002}),
  \eprint{arXiv:hep-ph/0201195}.

\bibitem[{\citenamefont{{ATLAS
  Collaboration}}(2012{\natexlab{b}})}]{ATLAS:2012ceu}
\bibinfo{author}{\bibnamefont{{ATLAS Collaboration}}},
  \bibinfo{journal}{ATLAS-CONF-2012-155, ATLAS-COM-CONF-2012-164}
  (\bibinfo{year}{2012}{\natexlab{b}}).

\end{thebibliography}

\end{document}